\documentclass[twocolumn]{aastex63}
\usepackage{amsmath}
\usepackage{wasysym}
\usepackage{comment}
\usepackage{threeparttable}
\usepackage{bm}       
\usepackage{braket}    
\accepted{July 15, 2023}
\shorttitle{Filament-Filament Collision}
\shortauthors{Kashiwagi, Iwasaki, and Tomisaka}
\graphicspath{{./}{figures/}}
\begin{document}

\title{Simulation of Head-on Collisions Between Filamentary Molecular Clouds Threaded by a Lateral Magnetic Field and Subsequent Evolution
}

\correspondingauthor{Raiga Kashiwagi}
\email{raiga.kashiwagi@grad.nao.ac.jp}

\author[0000-0002-1461-3866]{Raiga Kashiwagi}
\affiliation{Department of Astronomical Science, School of Physical Sciences, The Graduate University for Advanced Studies (SOKENDAI), 2-21-1 Osawa, Mitaka, Tokyo 181-8588,
Japan}

\author[0000-0002-2707-7548]{Kazunari Iwasaki}
\affiliation{Department of Astronomical Science, School of Physical Sciences, The Graduate University for Advanced Studies (SOKENDAI), 2-21-1 Osawa, Mitaka, Tokyo 181-8588,
Japan}
\affiliation{Center for Computational Astrophysics, National Astronomical Observatory of Japan, 2-21-1 Osawa, Mitaka, Tokyo 181-8588, Japan}
\affiliation{Division of Science, National Astronomical Observatory of Japan, 2-21-1 Osawa, Mitaka, Tokyo 181-8588, Japan}

\author[0000-0003-2726-0892]{Kohji Tomisaka}
\affiliation{Department of Astronomical Science, School of Physical Sciences, The Graduate University for Advanced Studies (SOKENDAI), 2-21-1 Osawa, Mitaka, Tokyo 181-8588, Japan}
\affiliation{Division of Science, National Astronomical Observatory of Japan, 2-21-1 Osawa, Mitaka, Tokyo 181-8588, Japan}



\begin{abstract}
Filamentary molecular clouds are regarded as the place where newborn stars are formed.
In particular, a hub region, a place where it appears as if several filaments are colliding, often indicates active star formation.
To understand the star formation in filament structures, we investigate the collisions between two filaments using two-dimensional magnetohydrodynamical simulations.
As a model of filaments, we assume that the filaments are in magnetohydrostatic equilibrium under a global magnetic field perpendicular to the filament axis.
We set two identical filaments with an infinite length and collided them with a zero-impact parameter (head-on).
When the two filaments collide while sharing the same magnetic flux, we found two types of evolution after a merged filament is formed: runaway radial collapse and stable oscillation with a finite amplitude. 
The condition for the radial collapse is independent of the collision velocity and is given by the total line mass of the two filaments exceeding the magnetically critical line mass for which no magnetohydrostatic solution exists.
The radial collapse proceeds in a self-similar manner, resulting in a unique distribution irrespective of the various initial line masses of the filament, as the collapse progresses.
When the total line mass is less massive than the magnetically critical line mass, the merged filament oscillates, and the density distribution is well-fitted by a magnetohydrostatic equilibrium solution.
The condition necessary for the radial collapse is also applicable to the collision whose direction is perpendicular to the global magnetic field.

\end{abstract}


\section{Introduction}\label{sec:intro}

The early phase of star formation can be understood by studying the filamentary structures in molecular clouds. 
Filaments are elongated structures within the dense parts of molecular clouds, and the Herschel space telescope showed that they are a fundamental component of these clouds \citep{2010A&A...518L.102A,2014prpl.conf...27A}.
Furthermore, prestellar and protostellar cores exist along such filaments \citep{2015A&A...584A..91K,2019A&A...621A..42A}. 
Star formation seems to occur in dense clumps, which have evolved from the filaments.
Thus, filaments play an important role as the birthplace of stars. 

Magnetic fields are also essential to understanding the early stages of star formation in various aspects, such as changing the character of interstellar turbulence \citep[e.g.,][]{2002ApJ...564..291C,2012ApJ...761..156F}, modifying the gas compression by shocks \citep[e.g.,][]{2013ApJ...774L..31I, 2022ApJ...934..174I}, achieving stability against gravitational instability \citep[e.g.,][]{1978PASJ...30..671N}, and transferring the angular momentum from a star forming gas \citep[e.g.,][]{1980ApJ...237..877M,2000ApJ...528L..41T}.
Orientation of the magnetic field is observed using near-infrared, far-infrared, and millimeter-wave polarizations. 
These polarizations are obtained from dust grains aligned with the interstellar magnetic field, that is, the background starlight in the near-infrared is polarized parallel to the magnetic field, whereas the thermal emission from the magnetically aligned dust in the far-infrared and millimeter wavelengths is polarized perpendicular to the magnetic field. 
Previous observations have indicated that the global magnetic field is nearly perpendicular to the massive filament \citep{2011ApJ...741...21C,2013A&A...550A..38P,2019PASJ...71S...7S,2020ApJ...899...28D,2021A&A...647A..78A}, and, from the Planck all-sky survey, 
Planck Collaboration Int.\,XXXV (\citeyear{2016A&A...586A.138P}) found that the interstellar magnetic field is often perpendicular to massive filaments but is parallel to filaments with a low column density (striations). 
This trend is evident in typical molecular clouds such as Taurus, Lupus, and Chamaeleon–Musca.

One of the formation scenarios of such filaments is explained by the colliding streams \citep{2015A&A...580A..49I}.
Colliding streams form a compressed layer, which is a flattened sheet-like structure.
The physical origin of this stream corresponds to the expanding H\textsc{ii} region, supernova remnant, or collision of two clouds.
In this context, the filament is formed by the self-gravitational fragmentation of the sheet-like clouds.

The self-gravitational instability of an equilibrium sheet is well-established. 
Perturbations with wavelengths longer than a critical value can grow, leading to the fragmentation of sheets into filaments.
The critical wavelength of a nonmagnetized sheet is typically a few times larger than its thickness. 
\citet{1987PThPh..78.1273M} performed a detailed analysis of the nonlinear growth of unstable perturbations and found that sheet-like clouds are expected to fragment into filaments separated by approximately twice the critical wavelength.
The presence of a perpendicular magnetic field tends to stabilize the sheet.
\citet{1978PASJ...30..671N} showed that if the magnetic field strength is larger than the critical value ($B_\mathrm{crit}=2\pi G^{1/2}\Sigma$, where $\Sigma$ is the column density of the sheet), the sheet is stable against gravitational fragmentation.
When the magnetic field is uniform and in the plane of the sheet, its stabilizing effect is limited \citep{1983PASJ...35..187T}, but the direction of the fragmentation is determined by the strength of the external pressure \citep{1998ApJ...506..306N}.
When the external pressure is considerably smaller than the central pressure of the sheet, filaments are formed in the perpendicular direction to the mean magnetic field lines and have a line mass $\lambda$ heavier than the maximum line mass of a filament supported only by the thermal pressure $\lambda_\mathrm{crit}$ (see Equation (\ref{eq:line_mass}) below for further details), as $\lambda\sim 2\lambda_\mathrm{crit}$. 
Such dense filaments have been observed using the Herschel space telescope as star-forming filaments \citep{2010A&A...518L.102A}.
By contrast, when the external pressure is comparable to the central pressure of the sheet, the resulting filament is parallel to the mean magnetic field, with a line mass being significantly smaller than the critical one ($\lambda\ll\lambda_\mathrm{crit}$).
This type of filament may be known as a ``striation'' observed using the Herschel telescope \citep{2013A&A...550A..38P}.

Recently, several observations have indicated that active star formation occurs at the junction or overlap point of the filaments \citep[e.g.,][]{2010A&A...518L.102A,2014ApJ...791L..23N,2015A&A...574L...6F,2019ApJ...884...84D,2019ApJ...886...15T,2020A&A...642A..87K}. 
For example, \citet{2014ApJ...791L..23N} suggested that the formation of a star cluster in the Aquila Serpens South region, which is one of the active star-forming regions nearby the molecular clouds, was triggered by the collision between three filaments.  
\citet{2020A&A...642A..87K} reported that, in high-mass star-forming regions, all luminous clumps ($L\ge10^4L_\odot$) exist in the hub-filamentary structures.
These structures are defined as the junction of the filaments and are believed to be formed because of the coalescence of massive filaments. 
Therefore, the hub-filamentary structures are essential to understanding the massive star formation.

Based on the findings of these observational studies, there is a need to perform a theoretical study of the induced star formation caused by the collision between the filaments.
Thus far, only a few theoretical (numerical) studies have been conducted.
\citet{2011A&A...528A..50D} reproduced the star cluster at the Serpens region through a collision of the cylindrical clouds. 
\citet{2021MNRAS.507.3486H} determined the conditions for filament collisions by comparing the contraction time of a filament with a finite length and the required time for the filaments to collide. 
However, these studies did not consider the magnetic field for simplicity.
Star formation is significantly affected by the magnetic field, for example, shock compression and stability against gravitational instability.
Hence, a magnetohydrodynamical simulation study of filament collisions is required.

From a theoretical standpoint, in this study, we investigated the conditions for star formation induced by the collision between the filaments, including the magnetic field and the subsequent evolution.

In this paper, we developed a numerical simulation for explaining the filament collision. 
The remainder of this paper is organized as follows:
In section \ref{sec:basic_theory}, we briefly summarized the basic theory of the filament. In section \ref{sec:method}, we introduced a model 
for our numerical calculations 
In section \ref{sec:result}, we detailed the simulation results obtained from the filament collision 
In section \ref{sec:discussion}, we discussed the condition of the radial collapse of the merged filament.
In section \ref{sec:summary}, we summarized the results and conclusions of this study

\section{The critical line mass of the filament}\label{sec:basic_theory}

The stability of the isolated filaments is important in considering the initial condition of star formation and is often expressed with the mass per unit length, that is, line mass.
The hydrostatic equilibrium state of a nonmagnetized infinite cylindrical isothermal cloud has a density distribution $\rho(r)$, which is given as follows \citet{1963AcA....13...30S,1964ApJ...140.1056O}:
\begin{equation}\label{eq:iso_density}
    \rho(r)=\rho_c\left(1+\frac{r^2}{8H^2}\right)^{-2},
\end{equation}
where $\rho_c$ is the central density of the cylinder and $H$ is the scale height, which is expressed as $H=c_s/\left(4\pi G \rho_c\right)^{1/2}$ using the isothermal sound speed $c_s$ and the gravitational constant $G$. 
By integrating Equation (\ref{eq:iso_density}) into the radius $r$ to infinity, the line mass of the filament reaches a constant value as follows:
\begin{equation}\label{eq:line_mass}
    \lambda_{\rm crit}=\int^{\infty}_{0}2\pi \rho r dr =\frac{2c_s^2}{G}.
\end{equation}

This constant value is called the “critical line mass” and is often used as a criterion that determines whether the star formation begins in the filament or not \citep{1987PThPh..77..635N,1992ApJ...388..392I,2010A&A...518L.102A}. 

Filaments with a larger line mass than $\lambda_\mathrm{crit}$ have no hydrostatic solution, and radial contraction is induced in such filaments owing to self-gravity. By contrast, less massive filaments than $\lambda_\mathrm{crit}$ have hydrostatic equilibrium configurations.

As mentioned in Section \ref{sec:intro},
previous observations have shown that the filaments tend to have a magnetic field perpendicular to the long axis.
The magnetic fields affect the stability of the filaments as follows:
when the isothermal filament is threaded by a lateral magnetic field, the critical line mass of the magnetized filament ($\lambda_{\rm crit,B}$) is obtained through numerical calculations performed by \citet{2014ApJ...785...24T} and is expressed as a function of the magnetic flux as, 
\begin{equation}\label{eq:magnetized_critical_line_mass}
    \lambda_{\rm crit,B}\simeq 0.24 \frac{\Phi}{G^{1/2}}+1.66\frac{c_s^2}{G},
\end{equation}
where $\Phi$ represents one-half of the magnetic flux threading a filament per unit length, which has a dimension of the magnetic flux density of $B$ times the scale length $L$, that is, $[\Phi]=[B][L]$
(see also \citet{2021ApJ...911..106K} for the equivalent empirical relation for a polytropic gas filament.)
This empirical formula indicates that the magnetized critical line mass increases linearly by the magnetic flux, and the filament supports a large line mass by the Lorentz force compared with the nonmagnetized filaments, $2c_s^2/G$.
\footnote{
In Equation (\ref{eq:magnetized_critical_line_mass}), when $\Phi\rightarrow 0$, the magnetized critical value is smaller than the nonmagnetized value ($\lambda_{\rm crit,B}<\lambda_{\rm crit}=2c_s^2/G$). 
This is because this equation was obtained using the least square method; however, in practice, the magnetized critical line masses at small $\Phi$ converge at the nonmagnetized critical line masses. 
}

This empirical formula (Equation (\ref{eq:magnetized_critical_line_mass})) was obtained by plotting the maximum line mass of the magnetohydrostatic equilibrium solutions for a given magnetic flux 
\begin{equation}\label{eq:magnetic_flux}
    \Phi=R_0\cdot B_0.
\end{equation}
A series of equilibrium solutions were obtained for filaments whose mass distribution against the magnetic flux was considered as a hypothetical filament with a uniform density $\rho_0$ and a radius $R_0$, which was threaded by a uniform magnetic field $B_0$ and immersed in an external pressure of $p_\mathrm{ext}$.
In this case, an equilibrium configuration was specified with the hypothetical density $\rho_0$ or its line mass $\lambda=\rho_0 \pi R_0^2$, $R_0$, and $B_0$.
As is mentioned in \citet{2014ApJ...785...24T}, the density of the hypothetical filament $\rho_0$ was substituted with the central density $\rho_c$, which appeared in the equilibrium filament.

Based on these results, we investigate the dynamical evolution of filament--filament collision.

\section{Methods and models}\label{sec:method}

\subsection{Basic Equations of Ideal MHD} 

In this study, we investigated a head-on collision of two identical filaments penetrated by a magnetic field such that they share the same magnetic flux.
To simulate the filament--filament collision, we solve the following ideal magnetohydrodynamic equations, which can be written in a conservative form as follows:
\begin{equation}\label{eq:mass_conservation}
   \frac{\partial \rho}{\partial t}+\nabla\cdot(\rho \textit{\textbf{v}})=0,
\end{equation}

\begin{equation}\label{eq:equation_of_motion}
   \frac{\partial \rho\textit{\textbf{v}}}{\partial t}+\nabla\cdot\left(\rho \textit{\textbf{v}}\textit{\textbf{v}} -\frac{\textit{\textbf{B}}\textit{\textbf{B}}}{4\pi}+P_{\rm tot}\right)=-\rho\nabla \psi,
\end{equation}    
   
\begin{equation}\label{eq:induction_equation}
   \frac{\partial \textit{\textbf{B}}}{\partial t}-\nabla\times\left(\textit{\textbf{v}}\times \textit{\textbf{B}}\right)=0,
\end{equation}   
where $\rho$ is the density, $\bm{v}$ is velocity, $\bm{B}$ is the magnetic field, $P_{\rm tot}$ is the total pressure, which is the sum of the gas pressure and the magnetic pressure and is written as,
\begin{equation}\label{eq:total_pressure}
   P_{\rm tot}=p_{\rm gas}+\frac{B^2}{8\pi},
\end{equation}, 
and $\psi$ is the gravitational potential obtained by solving Poisson's equation expressed in the following equation: 
\begin{equation}\label{eq:poisson's_equation}
   \Delta \psi=4\pi G\rho,
\end{equation}  
where $G$ is the gravitational constant.
We assumed that the isothermal filaments are 
confined within an isothermal hot ambient medium, 
leading to the presence of two gases with different 
temperature.
To reproduce this, we implemented a scalar field, in which
the equation of state is given as,
\begin{equation}\label{eq:equation_of_state}
   p_{\rm gas}=S \rho,
\end{equation}  
where $S$ is the scalar field proportional to the temperature, and we assumed that $S$ evolves according to the advection of gas as follows:
\begin{equation}
 \frac{\partial S}{\partial t}+(\bm{v}\cdot \mathbf{\nabla})S=0.
\end{equation}
Further details regarding the scalar field is described in the next section (see section \ref{sec:init_condition}).

To solve the basic equations, we used Athena++ \citep{2020ApJS..249....4S} with the following setting:
the time evolution of these magnetohydrodynamic equations was solved using the third-order Runge--Kutta time integrator scheme, which was proposed by \citet{2009JSCom.38..251G}, and the Riemann problem was solved using the HLLD method \citep{2005JCoPh.208..315M}.
For spatial construction, the piecewise linear method was applied to the characteristic variables.
The constrained transport method was applied for maintaining the initial $\nabla \cdot \bm{B}=0$ condition \citep{1988ApJ...332..659E,2008JCoPh.227.4123G}.
The multi-grid algorithm was used to solve Poisson's 
equation (Equation (\ref{eq:poisson's_equation})) and was implemented 
in Athena++ by \citet{2023arXiv230213903T}.

In addition, we modified the multi-grid module of our two-dimensional calculations 
because it was originally designed for three-dimensional simulations.
The Jeans criterion \citep{1997ApJ...489L.179T} 
was used to avoid unexpected numerical fragmentation. 
We stopped the calculations when the relation between the Jeans length $L_{\rm J}=\sqrt{\pi c_s^2/(G\rho)}$ and the grid size $\Delta x$ breaks the condition of $4\Delta x\leq L_{\rm J}$.

\subsection{Initial and Boundary Conditions}\label{sec:init_condition}

\begin{figure*}
    \centering
    \includegraphics[keepaspectratio,scale=0.4]{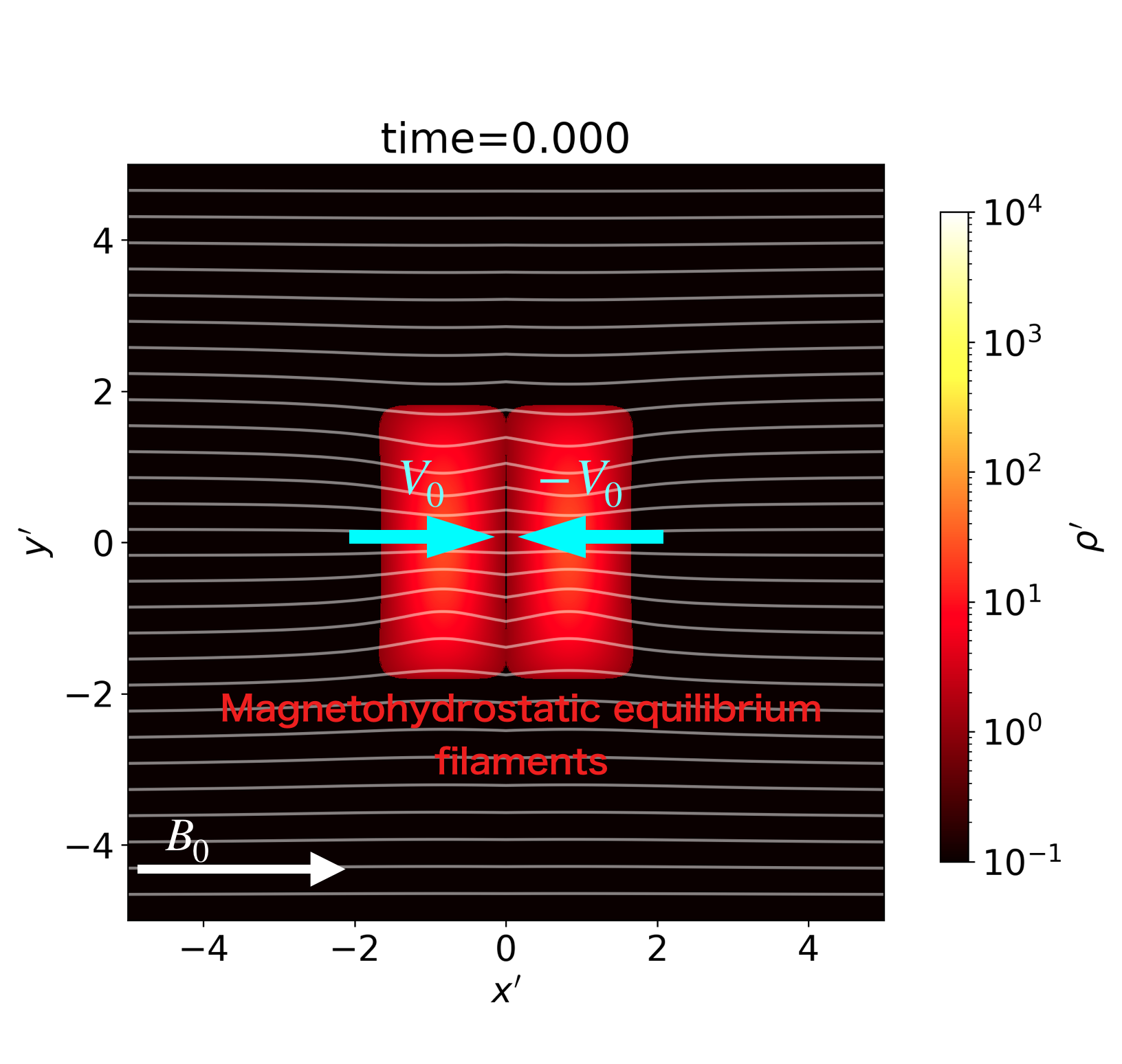}
      \caption{
      An example of the initial conditions: the density and magnetic field lines of the initial condition of the model $\beta01\mathrm{M}$ are plotted (see Table \ref{tab:tab1}).
      Two identical filaments in magnetohydrostatic equilibrium are placed close to each other.
      These filaments have initial velocities of $V_0$ and $-V_0$.
      White lines represent the magnetic field lines. 
      The external medium is shown in black.
      }
    \label{fig:scalar_field}
\end{figure*}

 We assume that the filaments are initially in magnetohydrostatic equilibrium threaded by the global magnetic field, which is perpendicular to their long axis. This magnetohydrostatic equilibrium solution is obtained by numerical calculations \citep{2014ApJ...785...24T,2021ApJ...911..106K}.
We then consider the initial condition of two colliding filaments, which are formed through fragmentation from the sheet, in which they shared the same magnetic flux \citep{1998ApJ...506..306N}.
The filaments are assumed to extend infinitely and uniformly in the $z$-axis direction in the Cartesian coordinate system $(x,y,z)$.
In the $x-y$ plane, we assume a computational domain of $-2.5 R_0 \le x, y \le 2.5 R_0$, where $R_0$ is the radius of the hypothetical parent cloud of the filament in magnetohydrostatic equilibrium (see Equation (\ref{eq:magnetic_flux})).
The magnetic field outside the filament is considered to be force-free and converged to the uniform magnetic field of $B_0$ with an increase in the distance from the filament $r\gg R_0$, following \citet{2014ApJ...785...24T}.
Figure \ref{fig:scalar_field} shows one of the initial states assumed in this study. 
The two identical filaments are adjacent to each other and moving in $\pm x$-directions.
The global magnetic field is parallel to the $x$-axis.

 In our simulation, we assumed that the gas temperature in the molecular cloud is constant, $T\simeq 5-10\,\mathrm{K}$; thus, the equation of state is isothermal. 
 However, we consider the possibility that the temperature of the gas may depend on its origin (filament or external medium). 
 Therefore, we use the scalar field (Equation (\ref{eq:equation_of_state})) in the equation of state.
 Initially, we assume that the filament and the external medium achieved pressure balance at the surface as follows:
 \begin{equation}
     p_s= S_\mathrm{fil}\rho_s= S_\mathrm{ext}\rho_\mathrm{ext},
 \end{equation}
 where $p_s$ is the pressure at the filament surface; 
 $\rho_s$ and $\rho_\mathrm{ext}$ represent the density at the surface 
 and that of the external medium, respectively; and
 $S_\mathrm{fil}$ and $S_\mathrm{ext}$ denote values of
 scalar $S$ inside and outside of the filament, respectively.
 We set the density contrast of the external medium and the filament surface as $\rho_\mathrm{ext}/\rho_s= 0.01$.
 Therefore, the scalar $S$ of the external medium is $S_\mathrm{ext}=100c_s^2$ and that of inside the filament is $S_\mathrm{fil}=c_s^2$, where $c_s$ is the sound speed inside the filament.
 Finally, in our simulation, we maintain the value of the sound speed of the external medium at $S_\mathrm{ext}^{1/2}=10c_s$.
 In other words, as the density of the external medium is lower than that of 
 the filament, we assume that the temperature of the external medium is higher than the 
 filament, and the initial temperature is maintained after the simulations are started. 
 
 The boundary conditions of the hydrodynamic variables were imposed to be periodic in the $y$-direction and outflow in the $x$-direction.
The boundary condition of the gravitational potential was assumed to be the Dirichlet condition, and the potential is calculated using multipole expansion.

\subsection{Normalization}\label{sec:Normalization}

\begin{table*}
    \caption{Units used for normalization}          
    \label{tab:normalization}      
    \centering          
    \begin{tabular}{ll}
    \hline 
        Unit of pressure\dotfill &  External pressure, $p_{\rm ext}$ \\
        Unit of density\dotfill &  Density at the surface, $\rho_s=p_\mathrm{ext}/c_s^2$\\
        Unit of time\dotfill &  Free-fall time, $t_{\rm ff}=(4\pi G\rho_s)^{-1/2}$ \\ 
        Unit of speed\dotfill & Isothermal sound speed in the filament, $c_s$ \\
        Unit of magnetic field strength \dotfill & $B_u=(4\pi p_{\rm ext})^{1/2}$\\
        Unit of length\dotfill & $L=c_{s}t_{\rm ff}=c_{s}/(4\pi G\rho_s)^{1/2}$\\
    \hline                  
    \end{tabular}
    \end{table*}

In this study, physical variables are normalized using the following quantities: external pressure ($p_\mathrm{ext}$), density of the filament surface ($\rho_s$), and sound velocity in the filament ($c_s$).
The scale length $L$ is defined by the free-fall time $t_\mathrm{ff}$ at the filament’s surface density $\rho_s$ and the isothermal sound speed inside the filament as $L=c_s t_\mathrm{ff}=c_s/(4\pi G\rho_s)^{1/2}$.
The physical scales characterizing the system are given in Table \ref{tab:normalization}.
We define the normalized variables as follows: $\bm{v}'\equiv \bm{v}/c_s$, $p'\equiv p_\mathrm{gas}/p_\mathrm{ext}$, $\rho' \equiv \rho/\rho_s$, $\Phi'\equiv \Phi/(B_uL)$, $\bm{B}'\equiv \bm{B}/B_u$, $\lambda'\equiv \lambda/(\rho_sL^2)$, $x' \equiv x/L$, and $y' \equiv y/L$, where the prime represents the normalized variables.
The scalar field is normalized using the isothermal sound speed in the filament as $S'\equiv S/c_s^2$.

\subsection{Parameters}\label{sec:parameters}

In this study, we consider three parameters: 
the field strength $B_0$, 
the total line mass $\lambda_\mathrm{tot}$,
and the collision speed $V_\mathrm{int}$.
The former two characterize the 
magnetohydrostatic state of each of the 
filaments, 
while $V_\mathrm{int}$ specifies the condition of the collision.

The nondimensional form of the field strength $B_0$ is given by $B'_0=B_0/(4\pi p_\mathrm{ext})^{1/2}=\sqrt{2}\beta_0^{-1/2}$, where $\beta_0 = 8\pi p_\mathrm{ext}/B_0^2$ is the plasma beta of the external gas far from the 
filaments.
In this study, $\beta_0$ is used instead of $B_0$
to specify the field strength.
We examine three plasma beta values: $\beta_0=1$ (weak), $\beta_0=0.1$ (fiducial), and $\beta_0=0.01$ (strong).

The total line masses are determined based on the magnetic critical line mass, 
whose nondimensional form is given by 
\begin{equation}\label{eq:dimensionless_magnetized_critical_line_mass}
    \lambda'_{\rm crit,B}\simeq 3.04\Phi'+20.8,
\end{equation}
where the nondimensional magnetic flux 
$\Phi' = R_0'\sqrt{2/\beta_0}$ 
is a function of $R_0'$ and $\beta_0$, that is, $\lambda_\mathrm{tot}'/2 \le \lambda_\mathrm{crit,B}'$.
For simplicity, the radius of the parental cylinder 
$R_0'$ is set to $2$.
The magnetic critical line masses for the three
$\beta_0$ are $\lambda'_{\rm crit,B}(\beta_0=1)=
1.17\lambda'_{\rm crit}$,
$\lambda'_{\rm crit,B}(\beta_0=0.1)=
1.91\lambda'_{\rm crit}$, and 
$\lambda'_{\rm crit,B}(\beta_0=0.01)=
4.25\lambda'_{\rm crit}$,
where 
$\lambda'_\mathrm{crit}$ is the 
nondimensional form of the 
critical line mass of Equation (\ref{eq:line_mass}) 
as $\lambda'_\mathrm{crit}=\lambda_\mathrm{crit}/(\rho_sL^2)=8\pi$.
We examine three types of line masses for the respective magnetic strengths, in which the filament in models L (massive), M (intermediate), and S (less massive) has a line mass of $0.9$, $0.75$, and $0.5$ times as large as the magnetically critical line mass, $\lambda_\mathrm{crit,B}'(\beta_0)$, respectively.
We also calculated additional models of S' (0.6$\lambda_\mathrm{crit,B}'(\beta_0)$) and M' (0.7$\lambda_\mathrm{crit,B}'(\beta_0)$), if necessary.

For calculating the collision speed $V_\mathrm{int}$, 
the cases with $V_\mathrm{int}=c_\mathrm{s}$ and 
$10c_\mathrm{s}$ are considered.

In short, the head-on collision was reproduced using the plasma beta value ($\beta_0$), total line mass ($\lambda'_\mathrm{tot}$), and initial relative velocity ($V'_{\rm int}$) as the parameters. 
In this study, we simulate 16 different models, shown in Table
\ref{tab:tab1}.
Hereafter, we omit $'$ with normalized variables, unless the quantities are misunderstood as dimensional variables.

\begin{table*}
\caption{Model parameters}
\label{tab:tab1}  
\begin{threeparttable}
\centering         
\begin{tabular}{lccccccc} 
\hline\hline
  $\mathrm{Model}^{(1)}$& ${\beta_0}^{(2)}$ & ${\lambda_{\rm tot}/\lambda_{\rm crit}}^{(3)}$ & ${V_{\rm int}/c_s}^{(4)}$  & ${\rho_{c}/\rho_s}^{(5)}$& ${\rho_{\rm lim}/\rho_s}^{(6)}$ &${N_{\rm mesh}}^{(7)}$ & $\mathrm{Result}^{(8)}$\\ \hline
    $\beta1$S  & 1& 1.17  & 1 & 3.30 & $4.13\times10^5$& $4096\times4096$ & stable\\
    $\beta1$S' & 1& 1.40  & 1 & 5.00 & $1.65\times10^6$ & $8192\times8192$ & radial collapse\\
    $\beta1$M & 1& 1.76  & 1 & 11.0 & $1.65\times10^6$ & $8192\times8192$ & radial collapse\\
    $\beta1$L & 1&  2.11   & 1 & 46.90 & $1.65\times10^6$ & $8192\times8192$ & radial collapse\\
    $\beta01$S  & 0.1& 1.91  & 1 & 6.80 & $4.13\times10^5$ & $4096\times4096$ & stable\\
    $\beta01$S\_{highV}  & 0.1& 1.91  & 10 & 6.80 & $4.13\times10^5$ & $4096\times4096$ & stable\\
    $\beta01$S' & 0.1 & 2.29 & 1 & 10.90 & $1.65\times10^6$ & $8192\times8192$ & radial collapse\\
    $\beta01$M & 0.1 & 2.87 & 1 & 24.50 & $1.65\times10^6$ & $8192\times8192$ & radial collapse\\
    $\beta01$M\_{highV} & 0.1& 2.87  & 10 & 24.50 & $1.65\times10^6$ & $8192\times8192$ & radial collapse\\
    $\beta01$L & 0.1& 3.44  & 1 & 81.0 & $1.65\times10^6$ & $8192\times8192$ & radial collapse\\
    $\beta001$S & 0.01& 4.25  & 1 & 37.0 & $4.13\times10^5$ & $4096\times4096$ & stable\\
    $\beta001$S' & 0.01& 5.20  & 1 & 62.50 & $1.65\times10^6$ & $8192\times8192$ & radial collapse\\
    $\beta001$M & 0.01& 6.38  & 1 & 140.0 & $1.65\times10^6$ & $8192\times8192$ & radial collapse\\
    $\beta001$L & 0.01& 7.65  & 1 & 406.2 & $1.65\times10^6$ & $8192\times8192$ & radial collapse\\
    $\beta1$S'$\perp$ & 1& 1.40  & 1 & 5.0 & $1.03\times10^5$ & $2048\times2048$ & stable\\
    $\beta1$M'$\perp$ & 1& 1.64  & 1 & 8.20 & $1.03\times10^5$ & $2048\times2048$ & radial collapse\\
    \hline
\end{tabular}
\begin{tablenotes}
      \small
      \item {\bf Notes.} The table columns are as follows: (1) Name of the model, (2) plasma beta value,  (3) the total line mass normalized by the thermal critical line mass, (4) initial relative velocity, (5) central density of the initial filament, (6) maximum density due to numerical restriction, which is determined by the Jeans criterion that is $\rho'_\mathrm{lim}\simeq 1.66\times 10^6(L'_\mathrm{box}/10)^{-2}(N_\mathrm{grid}/8192)^2$, when the numerical box size $L'_\mathrm{box}=10$ is divided into $N_\mathrm{grid}\times N_\mathrm{grid}=8192\times 8192$ grid points, (7) number of grid points, and (8) dynamical state of the merged filament.
      In all the models, the numerical box size is set to $L'_\mathrm{box}=10$.
\end{tablenotes}
\end{threeparttable}
\end{table*}

\section{Results}\label{sec:result}
As shown in Table \ref{tab:tab1}, our results showed that the outcome of filament collisions can be divided into two categories: a radial collapse model and a stable model.
The radial collapse model is characterized by a monotonic increase in the central density of the merged filament and indicates the global collapse of the system.
For the stable models, the merged filament does not collapse globally but oscillates around the magneto-hydrostatic state. 
We focus on the typical evolution of the models by explaining the radial collapse model, first.

\subsection{Radial Collapse Model}\label{sec:collapse}

\begin{figure*}
    \centering
     \begin{tabular}{cc}
         \includegraphics[keepaspectratio,scale=0.4]{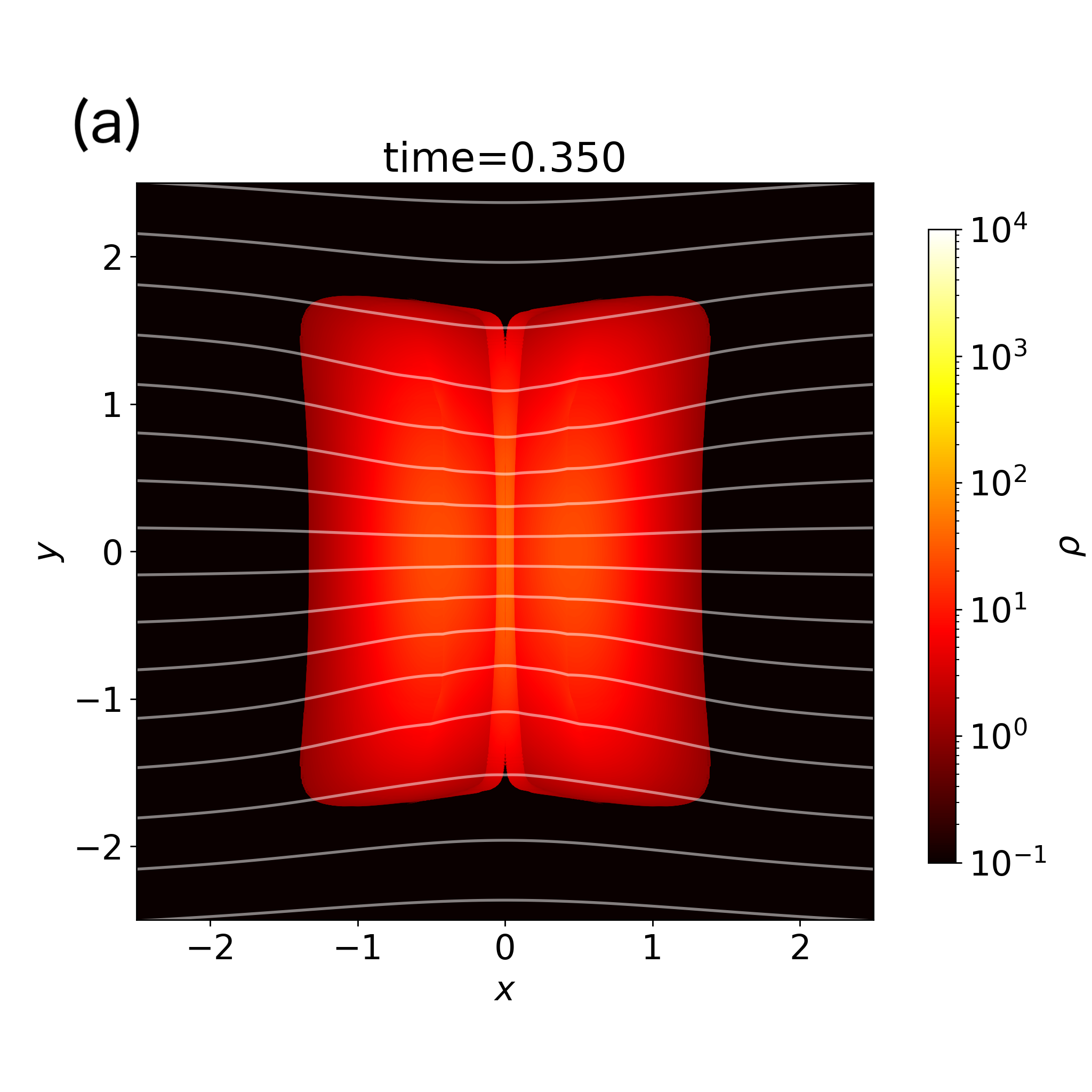} & \includegraphics[keepaspectratio,scale=0.4]{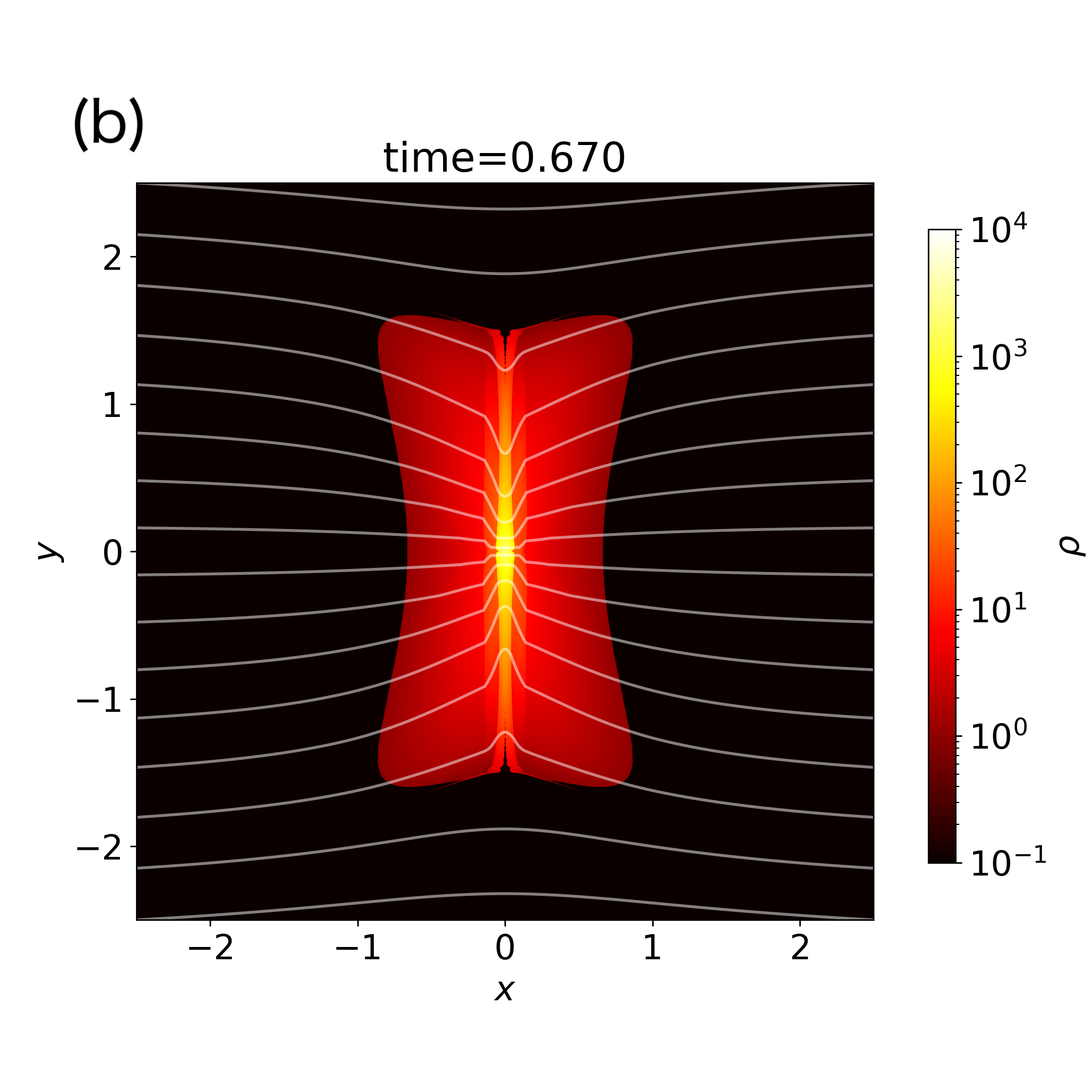}\\ \includegraphics[keepaspectratio,scale=0.4]{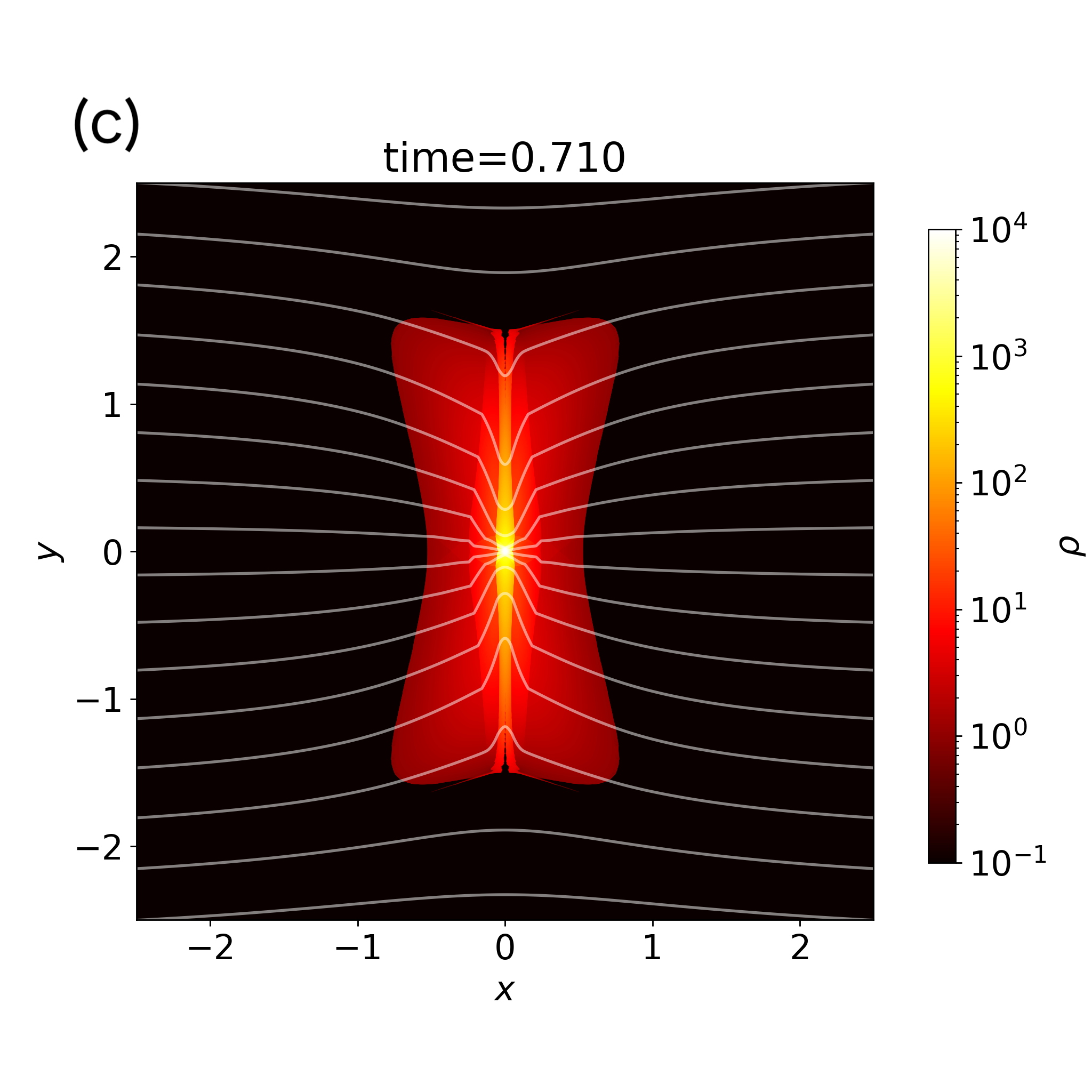}& \includegraphics[keepaspectratio,scale=0.4]{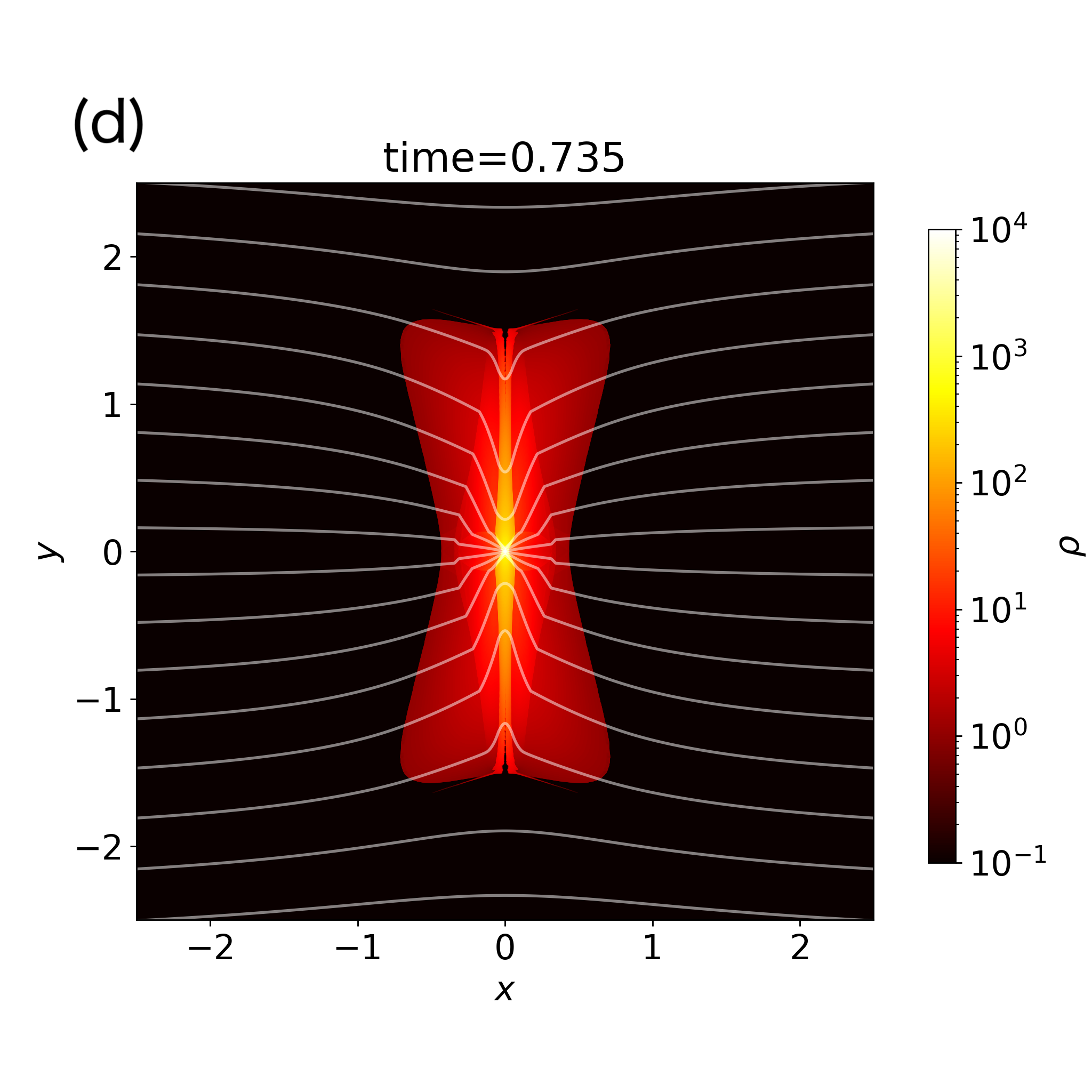}\\
     \end{tabular}
      \caption{
       Time evolution of the radial collapse model $\beta01$M. 
       This figure shows the cross-section of the colliding filaments on the $x-y$ plane, and the plot area is restricted to $\pm2.5$ from the box size ($\pm5$).
       Each panel shows the cross-section at different epochs: $t=0.350$ (\textit{a}), $0.670$ (\textit{b}), $0.710$ (\textit{c}), and $0.735$ (\textit{d}). 
       The color scale represents the density, and the white lines are the magnetic field lines.
      }
     \label{fig:2d_L075}
\end{figure*} 

\begin{figure*}
    \centering
     \begin{tabular}{cc}
         \includegraphics[keepaspectratio,scale=0.4]{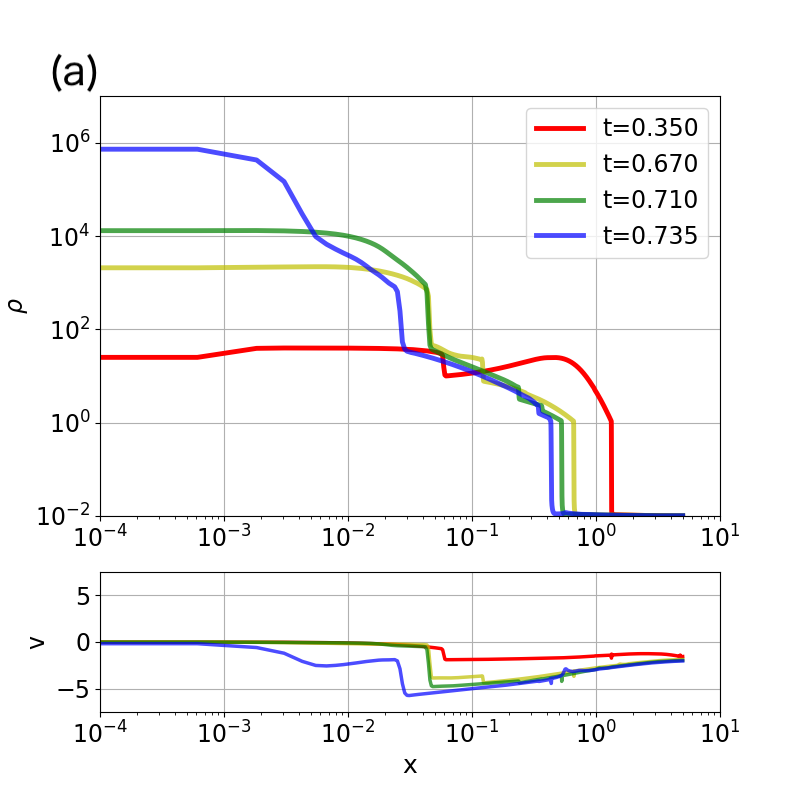}& \includegraphics[keepaspectratio,scale=0.4]{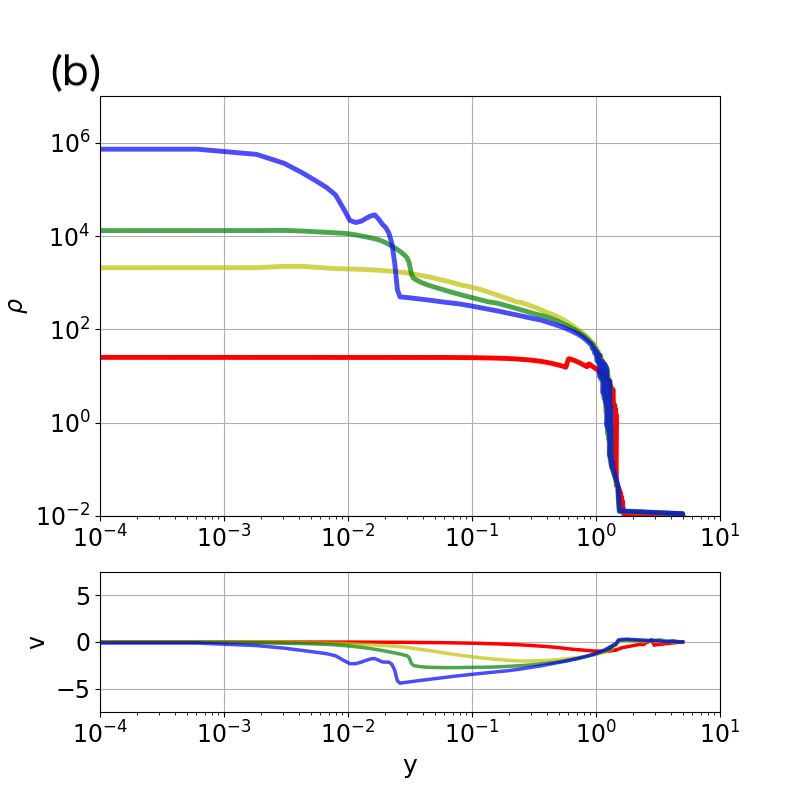}\\ \includegraphics[keepaspectratio,scale=0.4]{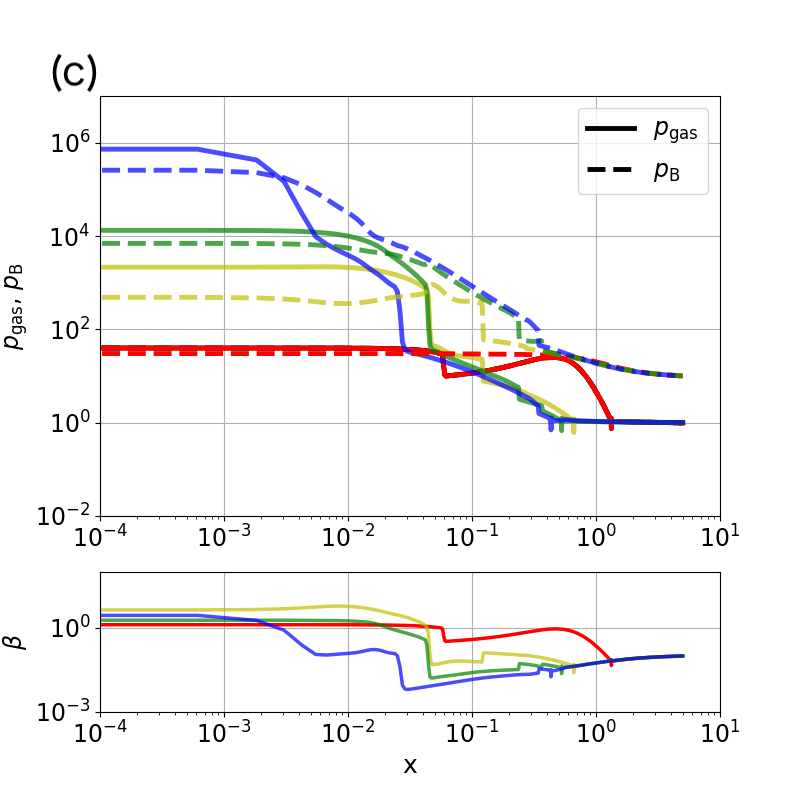}& \includegraphics[keepaspectratio,scale=0.4]{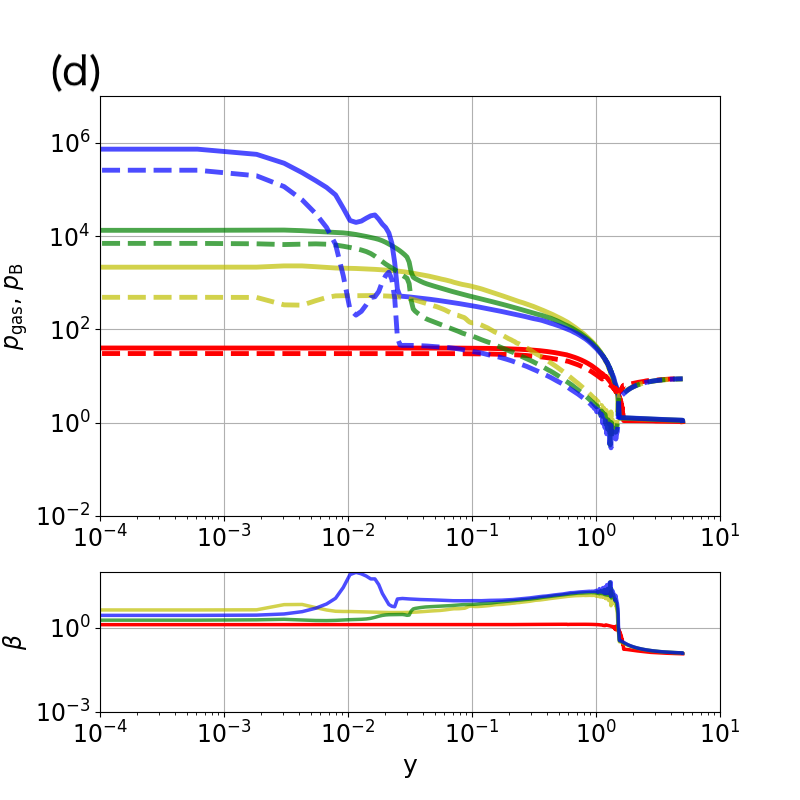}\\
     \end{tabular}
      \caption{
      The one dimentional profiles on the $x$- and $y$-axes for the radial collapse model $\beta$01M at four different epochs: $t=0.350$ (red), $0.670$ (yellow), $0.710$ (green), and $0.735$ (blue).
      The density and velocity profiles on the $x$- and $y$-axes are shown in (\textit{a}) and (\textit{b}), respectively,
      in which the solid lines represent 
      the density distributions ($\rho(x)$ in (\textit{a}) and $\rho(y)$ in (\textit{b})) in the upper panels,
      whereas these solid lines represent the velocity profiles ($v_x$ in (\textit{a}) and  $v_y$ in (\textit{b})) in the lower panels.
      The gas and magnetic pressure profiles on the $x$- and $y$-axes are shown in (\textit{c}) and (\textit{d}), respectively. 
      In (\textit{c}) and (\textit{d}), the solid and dashed lines in the upper panel represent the gas and magnetic pressure, respectively. 
      The solid lines in the lower panels represent the plasma beta values.
      }
    \label{fig:density_L075_new}
\end{figure*}

In this section, we describe the typical evolution of the radial collapse model. 
The model with $\beta_0=0.1$, $\lambda_\mathrm{tot}=2.87\lambda_{\rm crit}$, and $V_{\rm int}=1$ is selected as the fiducial model and corresponded to  $\beta01$M in Table \ref{tab:tab1}.

Figure \ref{fig:2d_L075} shows the images of the time evolution of the filament--filament collision.
In Figure \ref{fig:2d_L075}(\textit{a}), at $t=0.350$, the collision of two filaments resulted in the formation of a sheet-like dense structure, which is elongated in the $y$-direction and with a thickness of $\sim 0.1$ measured on the $x$-axis at the center of the merged filament, referred to as the shocked region.

Figures \ref{fig:density_L075_new}(\textit{a}) and (\textit{b}) show the density and velocity profiles on the $x$- and $y$-axes, respectively.
In the upper panel of Figure \ref{fig:density_L075_new}(\textit{a}), a shock front is visible at $|x|\simeq 0.06$ in the density profile.
This is visualized as the edge of the shocked region, which extends in the $y$-direction in Figure \ref{fig:2d_L075}(\textit{a}).
Along the $y$-axis, shock fronts have not yet formed at $t=0.350$ (see the upper panel of Figure \ref{fig:density_L075_new}(\textit{b})).
By comparing the density profiles on the $x$- and $y$-axes, 
we can observe that the density distribution of the merged filament in the $y$-direction ($|y|\simeq 1.5$) is wider than that in the $x$-direction ($|x|\simeq 0.06$).
In the lower panel of Figure \ref{fig:density_L075_new}(\textit{a}), the velocity distribution ($v_x$) at $t=0.350$ shows that a global inflow to the shocked region occurs in the $x$-direction outside the shocked region ($|x|\gtrsim 0.06$).
On the $y$-axis, the velocity distribution ($v_y$) does not show such a global inflow; however, an accretion flow is newly formed by the self-gravity near $|y|\simeq 1$ (see the lower panel of Figure \ref{fig:density_L075_new}(\textit{b})).

Figure \ref{fig:2d_L075}(\textit{b}) shows that the shocked region becomes denser at $t=0.670$ and is observed more clearly with time.
In addition, the magnetic field lines are dragged toward the center of the shocked region.
Magnetic field lines bend at the shock front reaching the front in the post-shock (inner) region, indicating that the shock wave created by the collision is a fast shock.

In the upper panel of Figure \ref{fig:density_L075_new}(\textit{a}), 
By comparing the density profiles at $t=0.350$ (red line) and $t=0.670$ (yellow line), 
We show that the overall density of the shocked region increases approximately by a factor of $100$. 

The position of the shock front moved from $|x|\simeq 0.06$ to $|x|\simeq 0.13$, indicating that the shock wave is expanding. 
The shock front, which is at $|x|\simeq 0.06$ at $t=0.350$, expands to $|x|\simeq 0.13$ 
at $t=0.670$.

Furthermore, another shock is observed at $|x|\simeq 0.04$, which continues to expand to the edge of the filament $|y|\lesssim 1.5$, as observed in Figure \ref{fig:density_L075_new}(\textit{b}).
The panel shows that the magnetic field lines bend at the front, departing from the front in the post-shock region $|x|\lesssim 0.04$. 
In addition, the lower panel of Figure \ref{fig:density_L075_new}(\textit{c}) shows that the plasma beta attains $\beta\simeq 0.13$ in the pre-shock region $|x|\gtrsim 0.1$, decreases with a jump passing the fast shock front at $|x|\simeq 0.13$, and increases with a positive jump at the accretion shock region $|x|\simeq 0.04$, which indicates that the second shock is a slow shock.
In other words, as a consequence of the supersonic collision of the filaments, a pair of outward-facing shock fronts are formed (outer fast and inner slow shocks).

On the contrary, along the $y$-axis, although the overall density of the shocked region increases with time, there is no clear evidence of the presence of a shock wave (see the upper panel of Figure \ref{fig:density_L075_new}(\textit{b})).
In the lower panel of Figure \ref{fig:density_L075_new}(\textit{b}), the velocity distribution ($v_y$) at $t=0.670$ shows that the central directional velocity component ranges from $|y|\simeq 10^{-2}$ to $|y|\simeq 2$. 
The lower panel of Figure \ref{fig:density_L075_new}(\textit{b}) shows that 
the gas in $10^{-2}\lesssim |y| \lesssim 2$ moves toward the center.
By comparing the velocity structures at $t=0.350$ and $t=0.670$, we show that the inflow is accelerated with time by the effect of self-gravity.

At $t=0.710$, Figure \ref{fig:2d_L075}(\textit{c}) shows that the fast shock front, which was formed by the collision, expands to $|x|\simeq 0.2$. 
Another slow shock is almost stalled around $|x|\simeq 0.04$. 
However, other differences from the previous epoch are not easily distinguished from the two-dimentional density profiles. 
Therefore, to more effectively analyze the evolution after the collision, we primarily focused on the radial profiles.
In the upper panel of Figures \ref{fig:density_L075_new}(\textit{a}) and (\textit{b}), the density profiles at $t=0.710$ (green lines) showed that the overall density of the shocked region 
continues to increase compared with the last epoch with $t=0.670$ (yellow). 
Although the slow shock facing toward the $x$-axis was almost stagnant near $|x|\simeq 0.04$, the density profile illustrated that the accretion shock is newly formed at $|y|\simeq 0.03$ along the $y$-axis.
This accretion shock is formed by the infalling gas as it reaches the high-density region, and the jump in the density and pressure at this location indicates the formation of an accretion shock.
This indicates that the gas is continuously accreted by the self-gravity of the shocked region, leading to an increase in the density and pressure of the region. 
The velocity profiles clearly show that the high-density shocked gas, even inside the accretion shock, contracts as a whole, primarily by the inflow in the $y$-direction
(see the lower panel of Figures \ref{fig:density_L075_new}(\textit{a}) and (\textit{b})).

At $t=0.735$, the 1D distribution of the merged filament near its final state is represented by the blue solid lines in Figures \ref{fig:density_L075_new}(\textit{a}) and (\textit{b}). 
The shock front contracts as a whole and reaches $|x|\simeq |y|\simeq 0.025$.
Inside $r\lesssim 5\times 10^{-3}$, the so-called “Hubble-like inflow” was observed as $v_x/x\simeq v_y/y\simeq \mathrm{const}$, and a uniform density core is contractiong in this region.

Next, we focus on the evolution of the magnetic field driven by the collision. 
Figures \ref{fig:density_L075_new}(\textit{c}) and (\textit{d}) show the time evolution of the gas ($p_{\rm gas}$), magnetic ($p_\mathrm{B}$) pressures, and plasma beta ($\beta\equiv p_{\rm gas}/p_\mathrm{B}$) along the $x$- and $y$-axes, respectively. 
The magnetic pressure is defined as $p_{\rm B}\equiv B^2/2$ in our nondimensional variables.
In the upper panels of Figures \ref{fig:density_L075_new}(\textit{c}) and (\textit{d}), we showed that the magnetic pressure within the shocked region is consistently smaller than the gas pressure throughout the evolution. 
As an example, at $t=0.735$, the plasma beta in the central part of the shocked region is $\beta=2.83$, meaning that, even in the presence of a magnetic field, thermal pressure dominates near the center of the shocked region.

The collision process leads to the amplification of the magnetic field strength, as observed by comparing the magnetic pressure of the central part at $t=0.350$ (red dashed line) and $t=0.735$ (blue dashed line) in the upper panel of Figures \ref{fig:density_L075_new}(\textit{c}) and (\textit{d}), which indicate that the magnetic pressure at the central part increases from $p_{\rm B}\simeq30$ to $p_{\rm B}\simeq2.6\times10^5$, and the amplification reaches a factor $\sim 10^4$. 

This suggests that the magnetic field strength is amplified by a factor of $\sim 10^2$ owing to the contraction induced by the collision.

\begin{figure}
    \centering
    \includegraphics[keepaspectratio,scale=0.4]{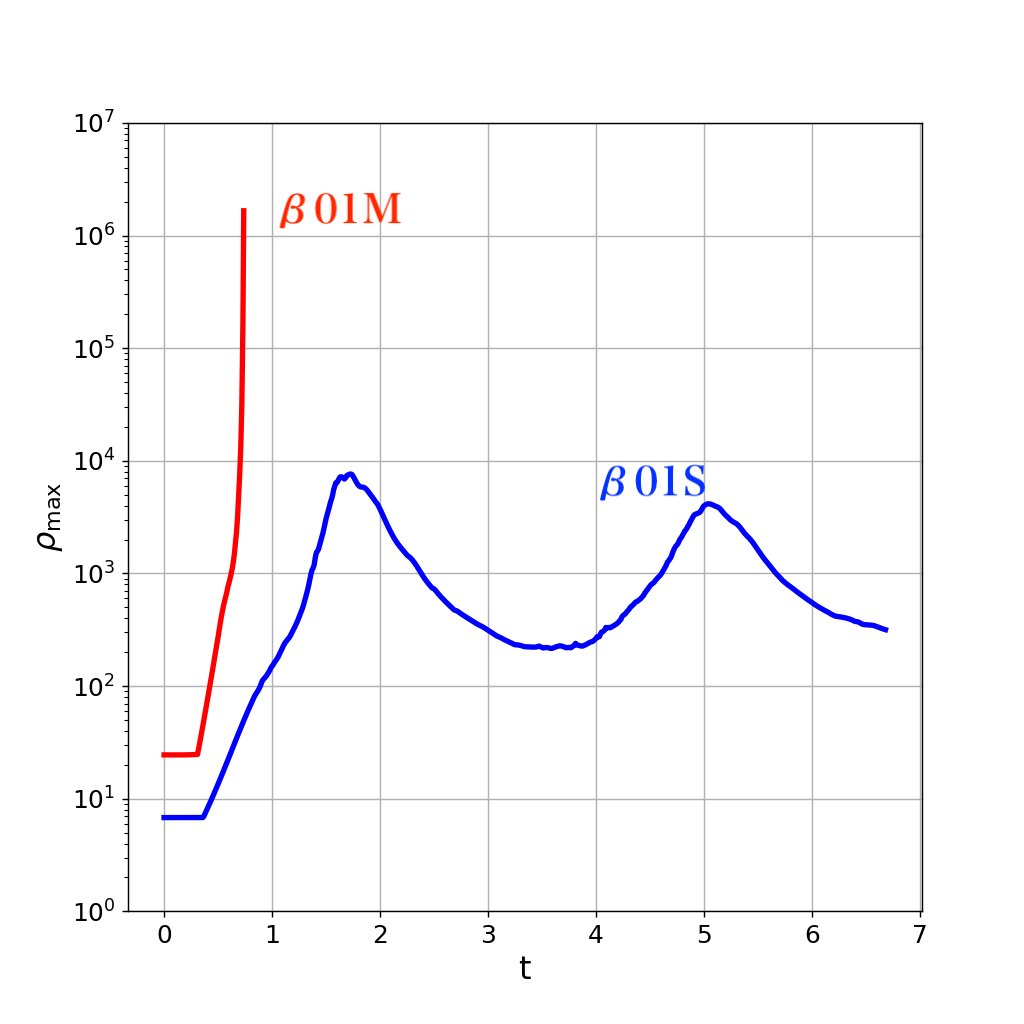}
      \caption{
      Maximum density $(\rho_\mathrm{max})$
      as a function of the elapsed time $(t)$.
      The red and blue lines represent the radial collapse ($\beta$01M) and stable ($\beta$01S) models, respectively.
      }
    \label{fig:maxden_L075_and_L05}
\end{figure}

In Figure \ref{fig:maxden_L075_and_L05}, we plot the evolution of maximum density $\rho_\mathrm{max}$ for the collapse ($\beta$01M) and the stable models ($\beta$01S).
Except for a special circumstance in the early phase of collision, maximum density is attained at the center of the merged filament as $\rho_\mathrm{max}=\rho_c\equiv\rho (0,0)$.
Figure \ref{fig:maxden_L075_and_L05} clearly shows that the central density increases monotonically with time, and the contraction never stops in the collapse model.

\subsection{Stable Models}\label{sec:stable}

\begin{figure*}
    \centering
     \begin{tabular}{cc}
         \includegraphics[keepaspectratio,scale=0.4]{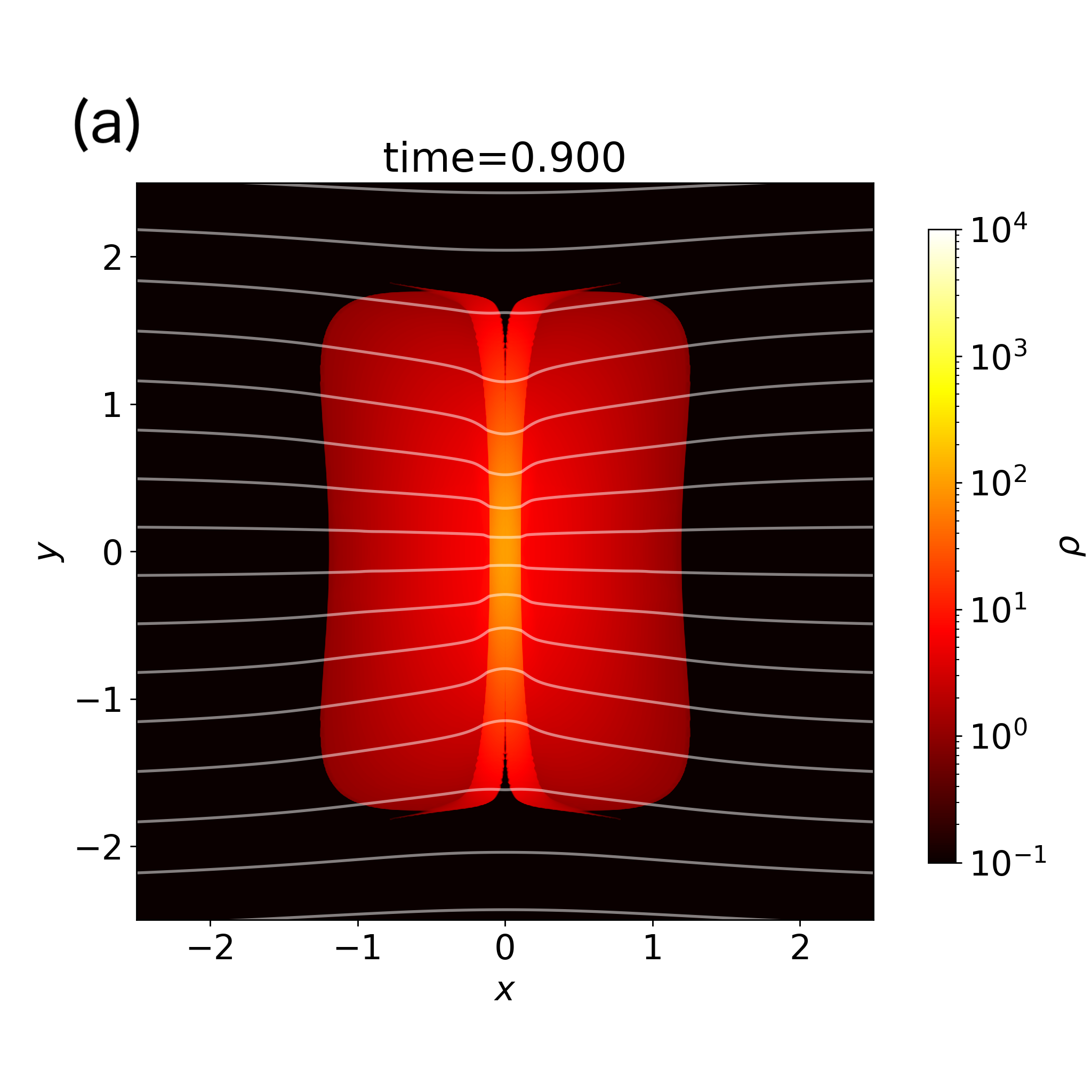} & \includegraphics[keepaspectratio,scale=0.4]{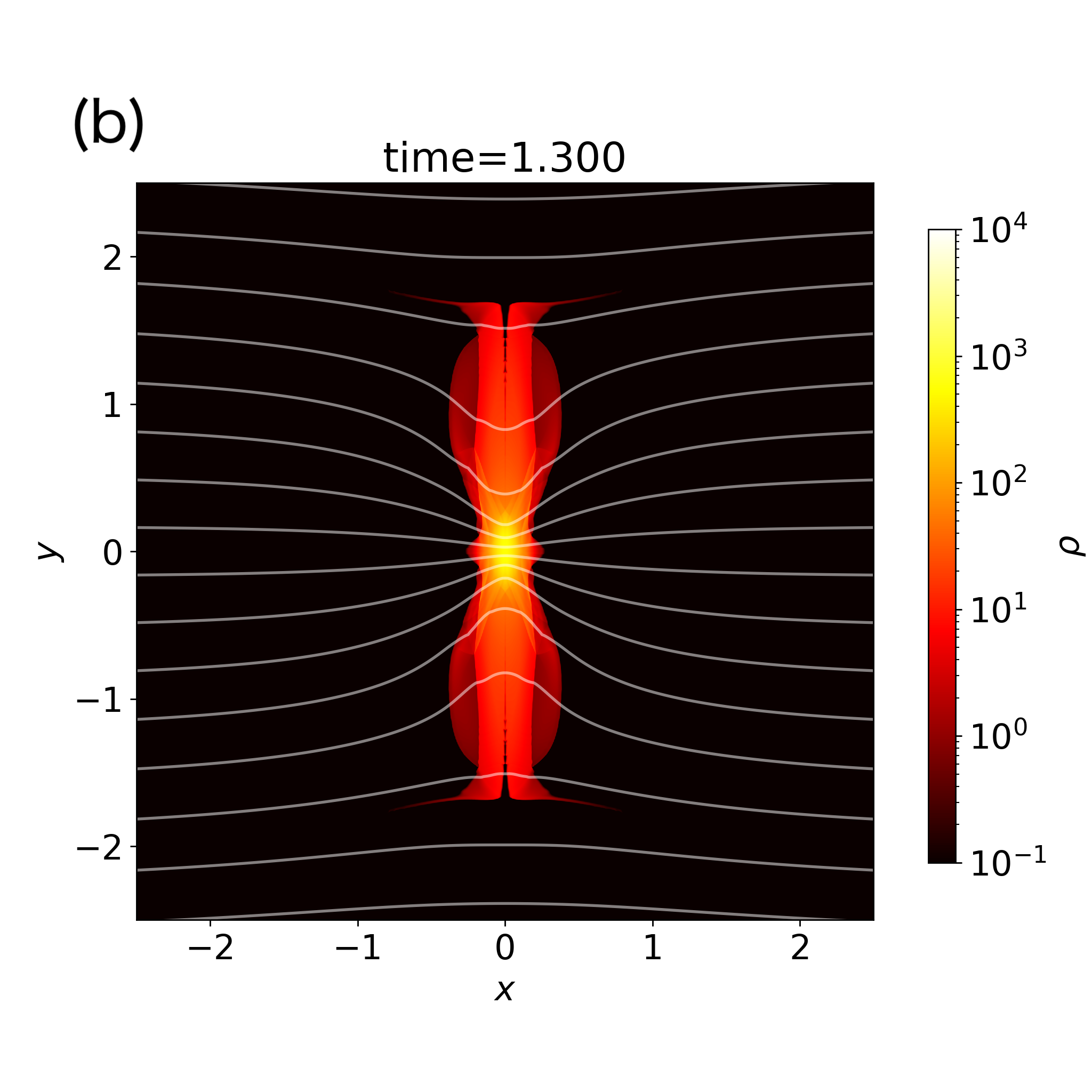}\\ \includegraphics[keepaspectratio,scale=0.4]{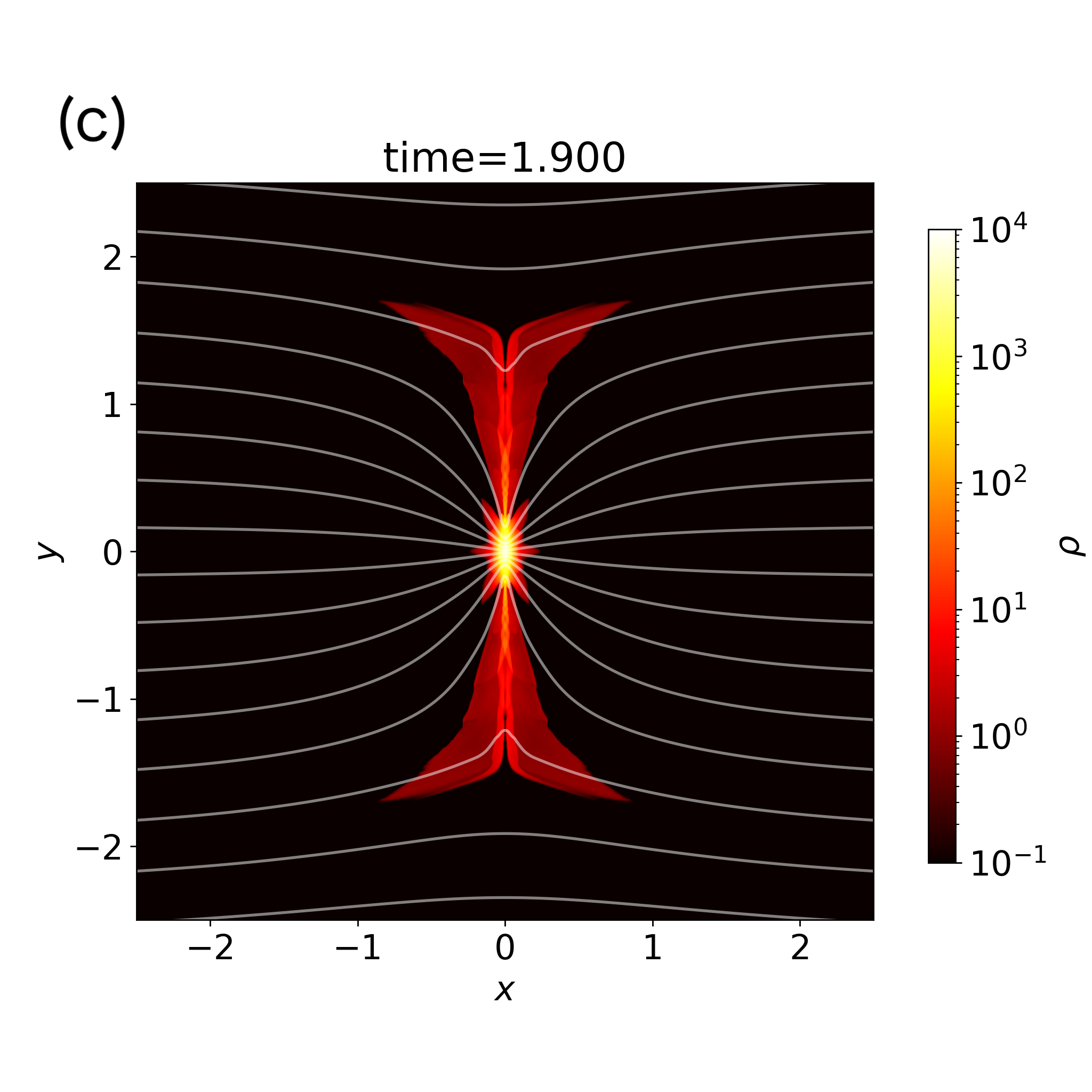}& \includegraphics[keepaspectratio,scale=0.4]{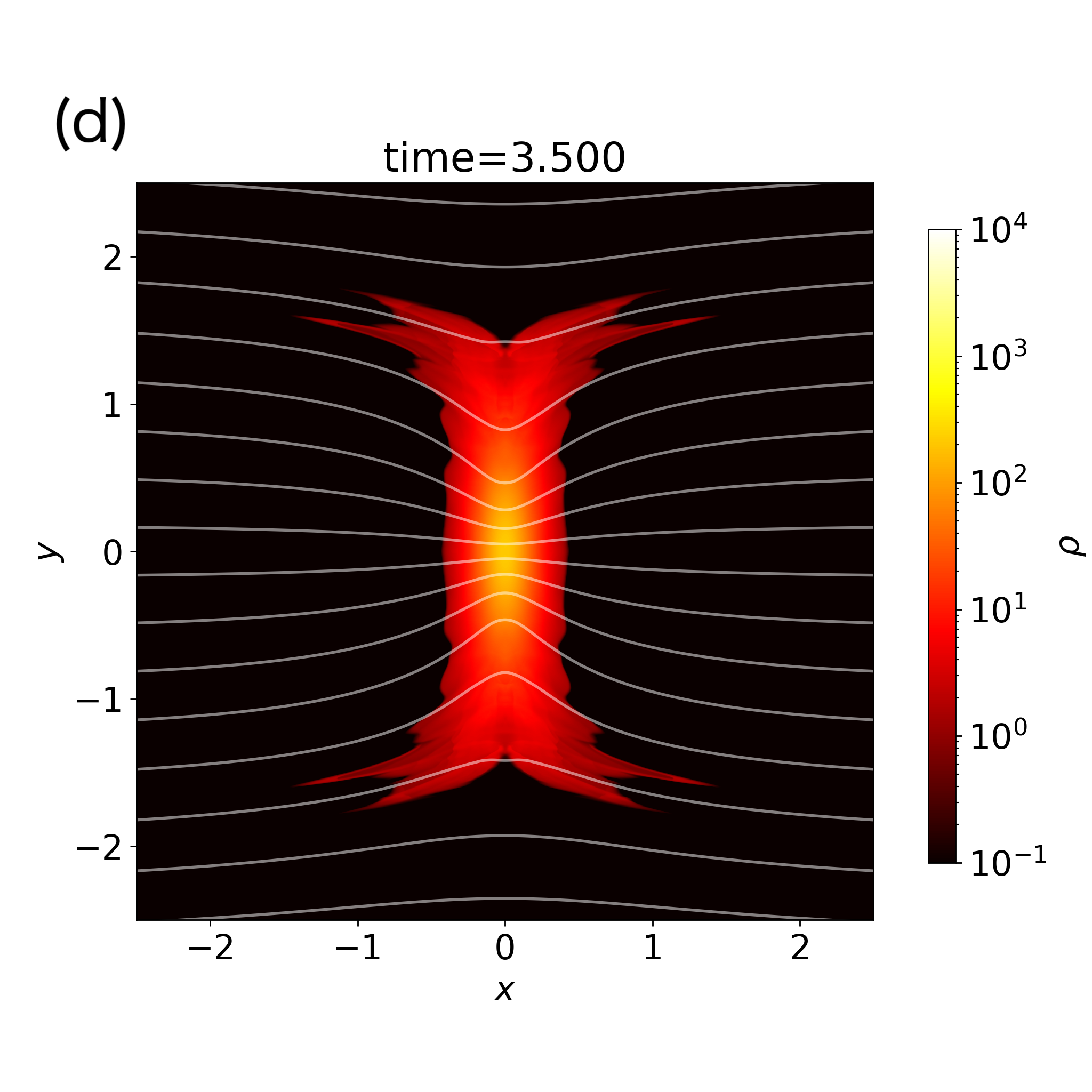}\\
     \end{tabular}
      \caption{
      Same as Figure \ref{fig:2d_L075}, but for the stable model ($\beta01$S).
      Each panel shows the cross-section at different epochs: $t=0.900$ (\textit{a}), $1.300$ (\textit{b}), $1.900$ (\textit{c}), and $3.500$ (\textit{d}). 
      }
     \label{fig:2d_L05}
\end{figure*}

\begin{figure*}
    \centering
     \begin{tabular}{cc}
         \includegraphics[keepaspectratio,scale=0.4]{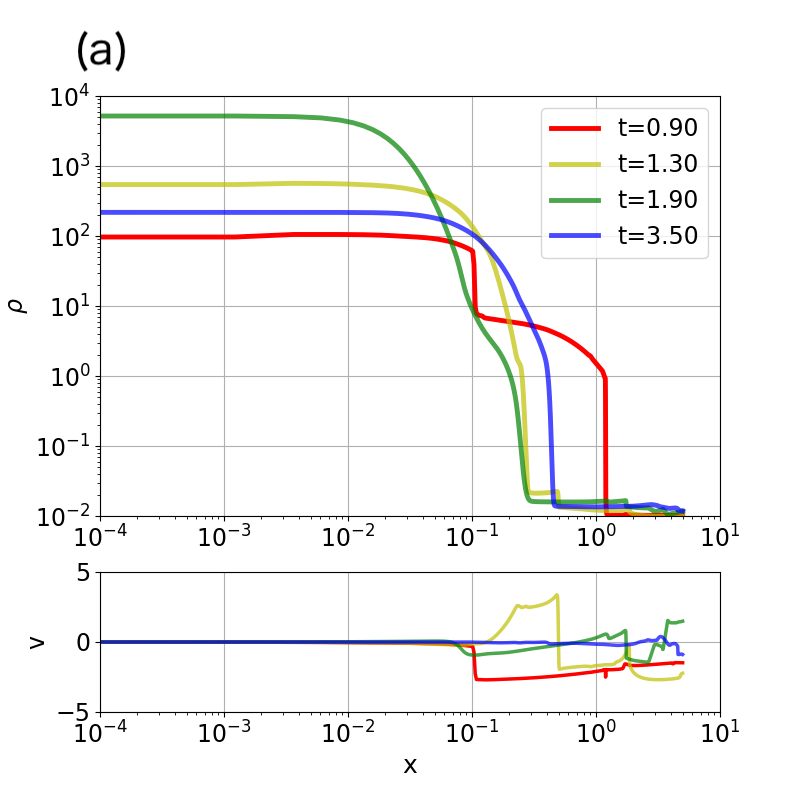} & \includegraphics[keepaspectratio,scale=0.4]{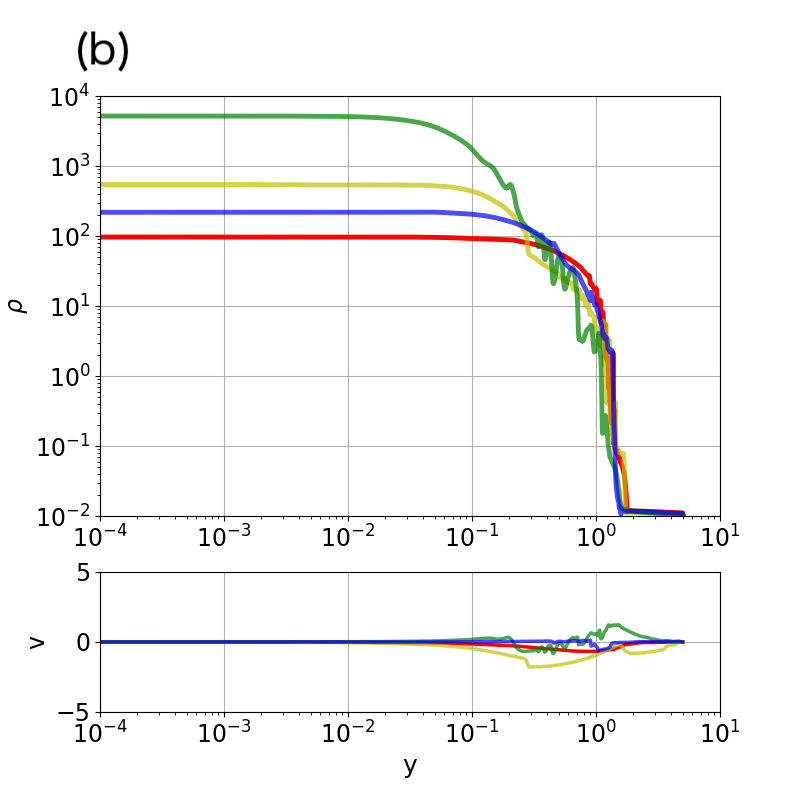}\\ 
     \end{tabular}
      \caption{
      Same as Figure \ref{fig:density_L075_new}, but for the stable model ($\beta01$S).
      Color represents different epochs: $t=0.900$ (red), $1.300$ (yellow), $1.900$ (green), and $3.500$ (blue).
      }
    \label{fig:density_L05_new}
\end{figure*} 

For stable runs, we develop the model $\beta01$S ($\beta_0=0.1$, $\lambda_\mathrm{tot}=1.91\lambda_\mathrm{crit}$, and $V_{\rm int}=1$).

The dynamical evolution of the stable models is similar to that of the radial collapse model during the stage at which the shock fronts are sweeping two colliding filaments.
Figure \ref{fig:2d_L05} shows four images of the time evolution after the collision of two filaments.
In Figure \ref{fig:2d_L05}(\textit{a}), at $t=0.900$, we can observe a sheet-like structure enclosed with two shock fronts facing outwardly, which is the same as that observed in the radial collapse model (Figure \ref{fig:2d_L075}(\textit{a})).
Figures \ref{fig:density_L05_new}(\textit{a}) and (\textit{b}) represent the density and velocity profiles along the $x$- and $y$-axes, respectively. 
The red lines correspond to the same epoch, $t=0.900$, as observed in Figure \ref{fig:2d_L05}(\textit{a}), which reveals that the shock wave is formed only along the $x$-axis ($|x|\simeq 0.1$). 
In the lower panel, velocity distributions (red lines) confirm this finding by showing that the global inflow to the shocked region is only observed along the $x$-axis.

At $t=1.300$, Figure \ref{fig:2d_L05}(\textit{b}) shows that the central part of the shocked region undergoes contraction and becomes denser, with the magnetic field lines also being dragged towards the center.
In Figure \ref{fig:density_L05_new}, the yellow lines indicate the density and velocity profiles at $t=1.300$. 
By comparing with the density profile of the previous epoch ($t=0.900$), we show that the central density of the shocked region increased by a factor of $\simeq6$. 
In addition, the density profile along the $y$-axis shows that a new accretion shock is formed at $|y|\simeq 0.3$.
In the lower panel of Figure \ref{fig:density_L05_new}(\textit{b}), the velocity distribution further supports this finding as $v_y$ indicates the presence of the accretion shock at $|y|\simeq0.3$, resulting from the global inflow occurring outside the new accretion shock ($|y|\gtrsim 0.3$). 
Early phase evolution is qualitatively the same as the collapsing model; for example, the formation of the shocked region initially occurred because of the $x$-axis global inflow (stage of $t=0.900$) and subsequently because of the $y$-axis accretion flow (stage of $t=1.300$).
However, in the stable model, the velocity profile on the $x$-axis demonstrated that the shock fronts observed in Figure \ref{fig:2d_L05}(\textit{a}) have swept across the merged filament, the boundary of which is observed at $|x|\simeq 0.50$, and are traveling in an external low-density medium (see the yellow line in the lower panel of Figure \ref{fig:density_L05_new}(\textit{a})).

In Figure \ref{fig:maxden_L075_and_L05}, we plotted the evolution of maximum density $\rho_\mathrm{max}(t)$ of this stable model in blue.
Figures \ref{fig:2d_L05}(\textit{c}) and (\textit{d}) depict the structures when $\rho_\mathrm{max}$ takes a maximum ($t\simeq 1.9$) and a minimum ($t\simeq 3.5$), respectively.
In Figure \ref{fig:density_L05_new},
a comparison between these density profiles at $t=1.9$ and $3.5$ shows that the density near the center decreases from $t=1.900$ to $t=3.500$, more than one order of magnitude; that is, by passing the maximum at $t=1.900$, the shocked region does not collapse but instead expands, as shown in Figure \ref{fig:2d_L05}(\textit{d}).
Furthermore, the magnetic field lines are dragged outward. 
After the expansion of the shocked region, the central density of the shocked region begins to increase again around $t \simeq 3.7$ and subsequently decreases again around $t \simeq 5.0$, as shown in Figure \ref{fig:maxden_L075_and_L05}.
Therefore, this stable model does not exhibit a global collapse but the shocked region continues to oscillate after the collision.

\begin{figure}
    \centering
    \includegraphics[keepaspectratio,scale=0.4]{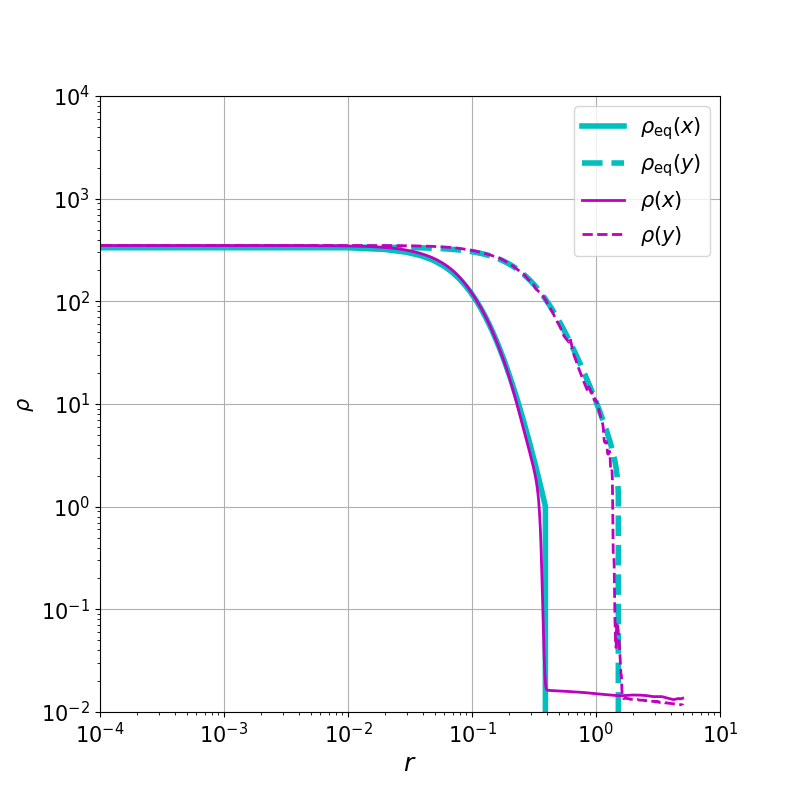}
      \caption{
      Density profiles on the $x$- and $y$-axes for the stable model
      $\beta01$S at time $t=6.500$, which corresponds to the epoch after the oscillation of two cycles.
      The magenta-colored solid and dashed lines correspond to the density profiles on the $x$- and $y$-axes, respectively.
      The cyan-colored solid ($\rho_{\rm eq}(x)$) and dashed ($\rho_{\rm eq}(y)$) lines correspond to the density profiles on the $x$- and $y$-axes, respectively, for the magnetohydrostatic equilibrium state \citep{2014ApJ...785...24T}.
      }
    \label{fig:density_L05B01}
\end{figure}

We then focus on the structure of the shocked region of the stable model.
Figure \ref{fig:density_L05B01} shows the density profile after oscillating for two cycles ($t=6.500$), in which the density profiles are plotted on the $x$- (magenta solid line) and $y$-axes (magenta dashed line). 
In this figure, we also plot the magnetohydrostatic equilibrium solution of \citet{2014ApJ...785...24T}, which has the same line mass and magnetic flux as the merged filament, represented in cyan.
Figure \ref{fig:density_L05B01} clearly indicates that the density profiles of the shocked region are quite similar to those of the magnetohydrostatic equilibrium state.

\subsection{Effect of Plasma Beta}\label{sec:beta}

\begin{figure*}
    \centering
     \begin{tabular}{cc}
         \includegraphics[keepaspectratio,scale=0.4]{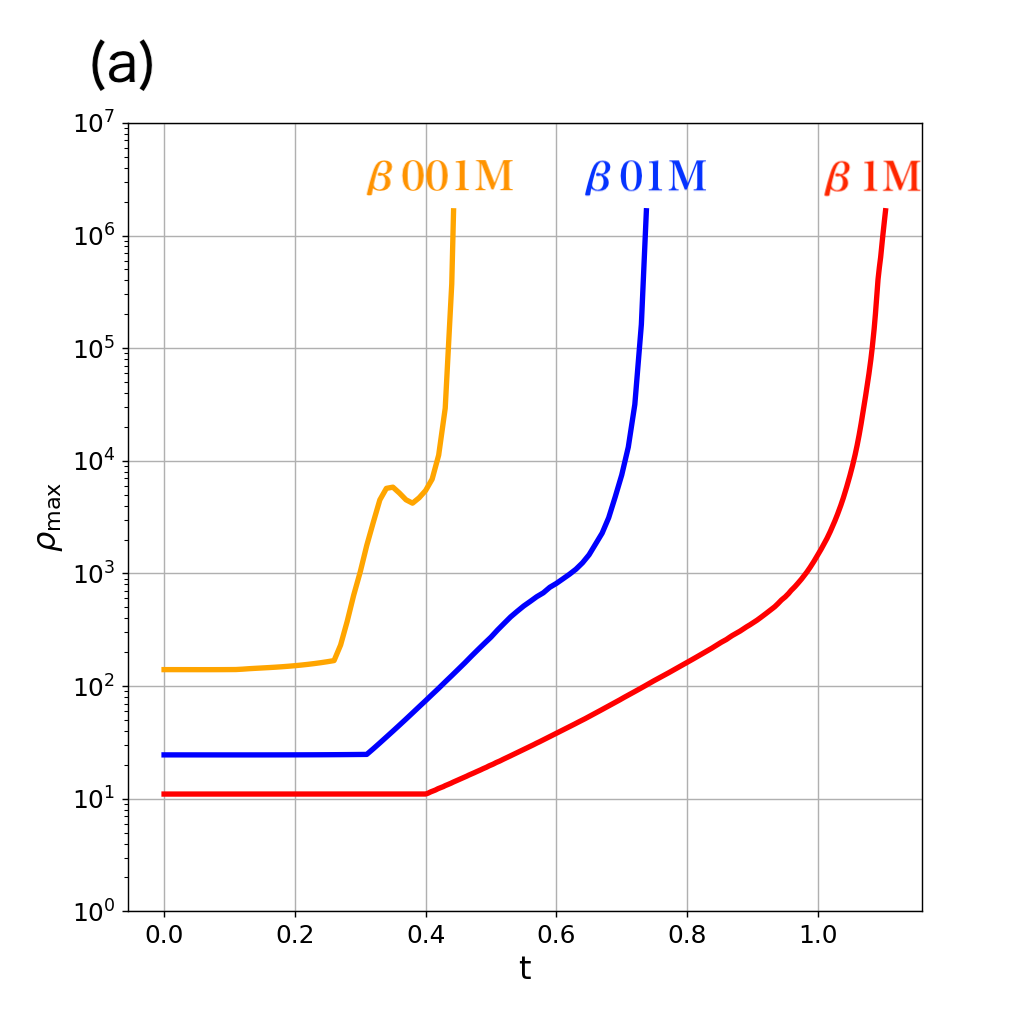} & \includegraphics[keepaspectratio,scale=0.4]{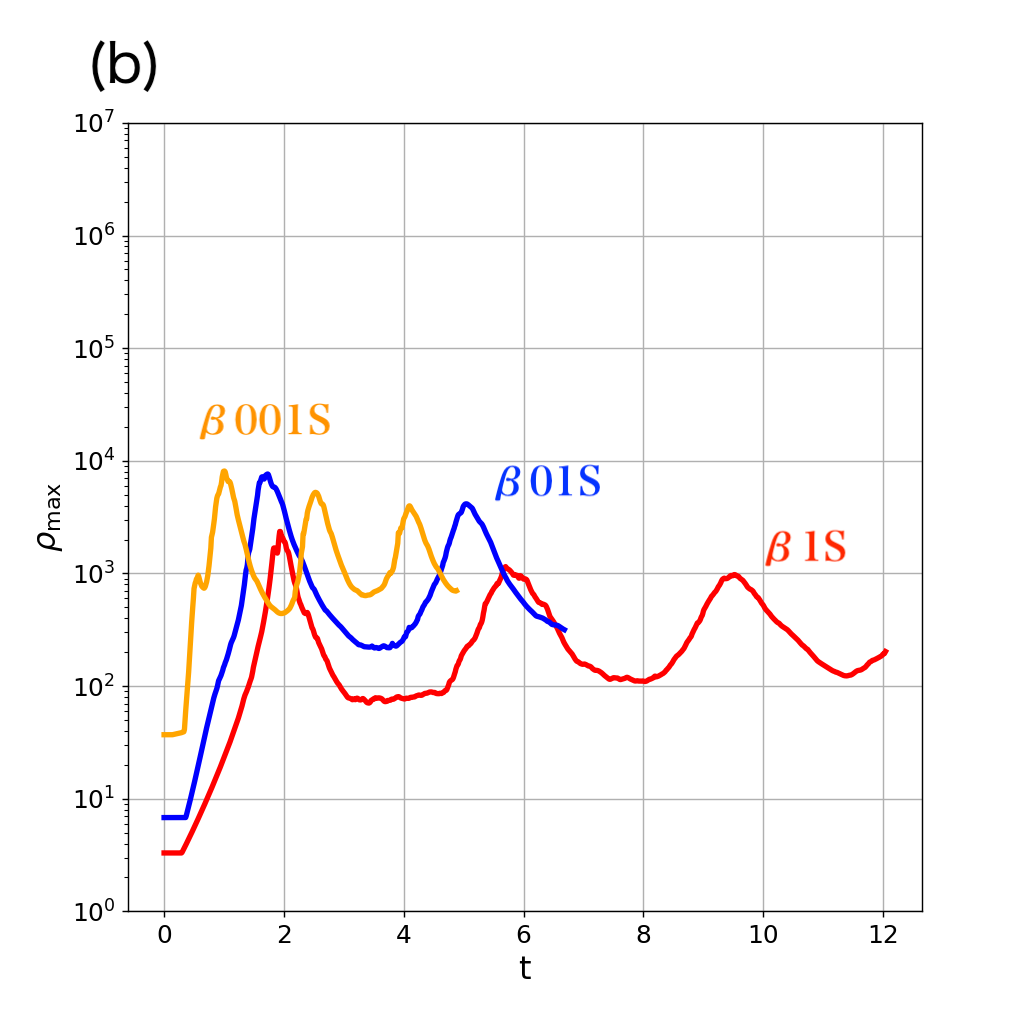}\\ 
     \end{tabular}
      \caption{ 
      Same as Figure \ref{fig:maxden_L075_and_L05}, but for the comparison of models with different $\beta_0$ values.
      Panels (\textit{a}) and (\textit{b}) show the maximum density evolution for the radial collapse and stable models, respectively. 
      Color represents different plasma beta values: $\beta_0=1$ (red), 0.1 (blue), and 0.01 (yellow).
      }
     \label{fig:beta_maxden}
\end{figure*}

\begin{figure*}
    \centering
     \begin{tabular}{cc}
         \includegraphics[keepaspectratio,scale=0.4]{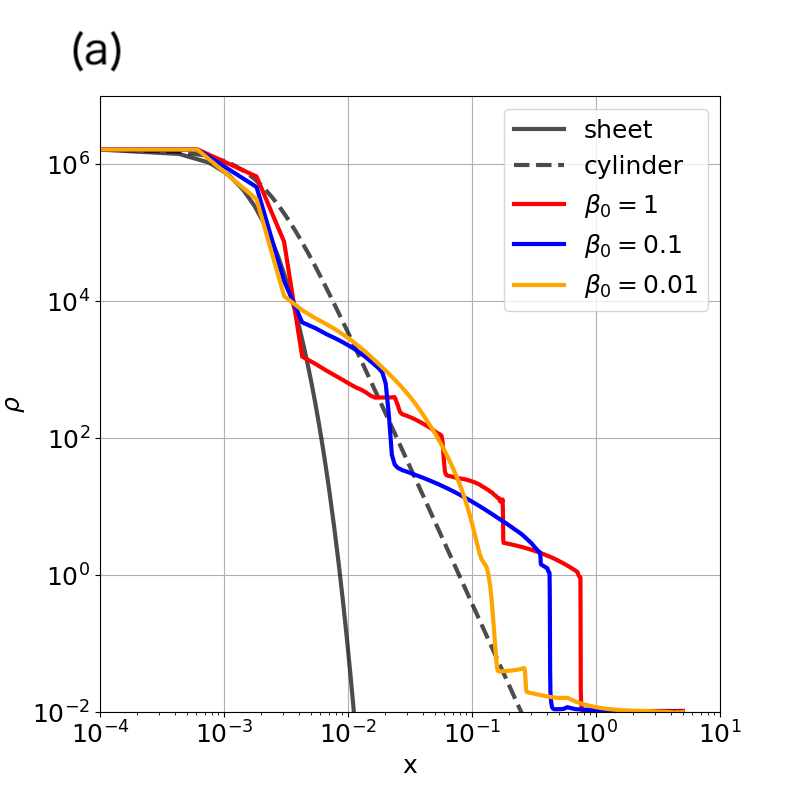} &  \includegraphics[keepaspectratio,scale=0.4]{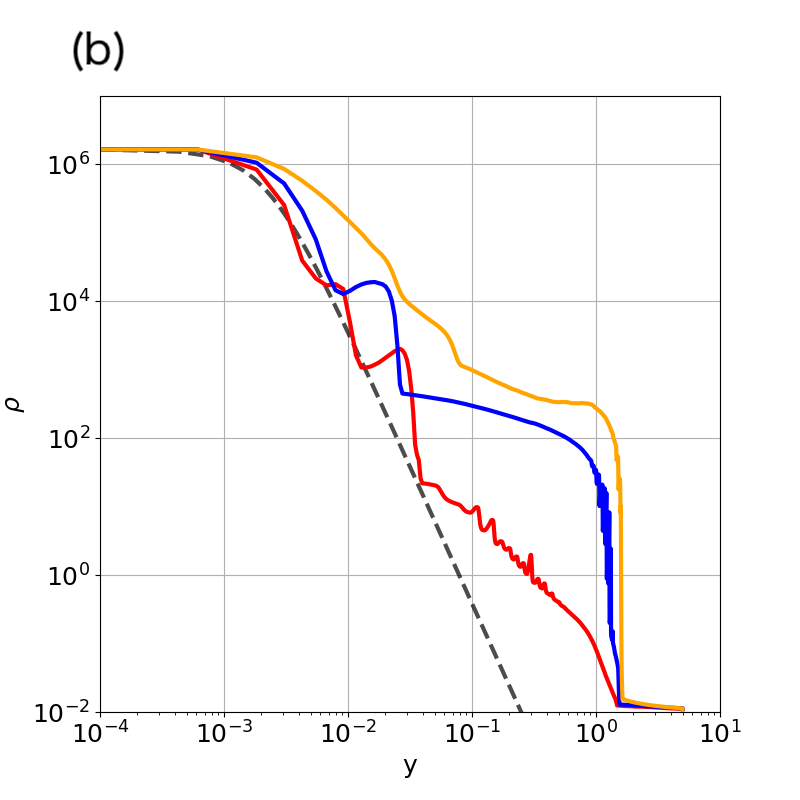}\\ 
     \end{tabular}
      \caption{
      The density profiles on the $x$- and $y$-axes for the radial collapse models with varying plasma beta values ($\beta_0$) are presented in panels (\textit{a}) and (\textit{b}), respectively.
      The colored solid lines show the results of different values of $\beta_0$: $\beta_0=1$ (red), 0.1 (yellow), and 0.01 (blue). 
      The black solid and dashed curves indicate the hydrostatic equilibrium solutions for the gas disk and filament, respectively.
      }
     \label{fig:L075_beta_density}
\end{figure*}

\begin{figure*}
    \centering
     \begin{tabular}{ccc}
         \includegraphics[keepaspectratio,scale=0.3]{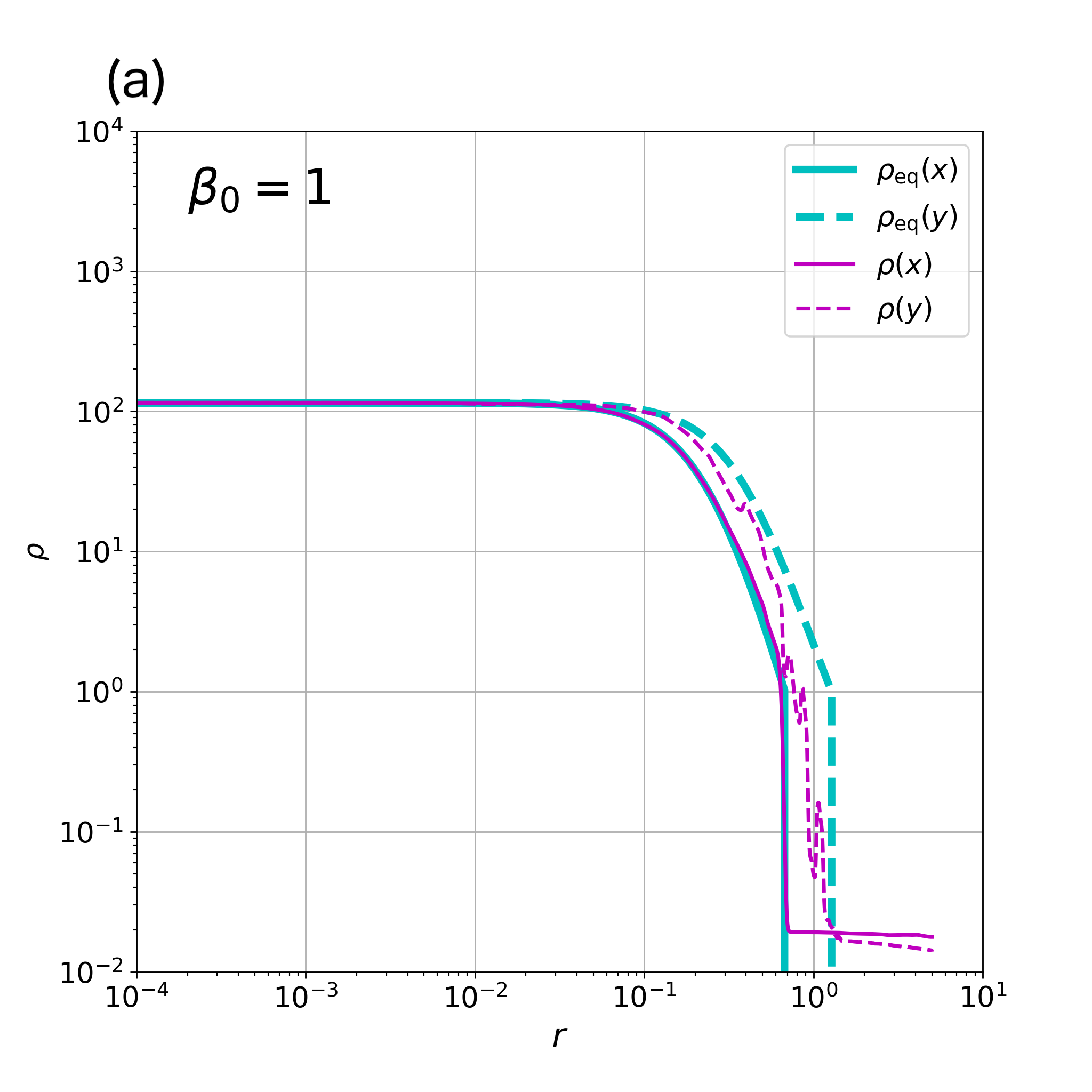} & \includegraphics[keepaspectratio,scale=0.3]{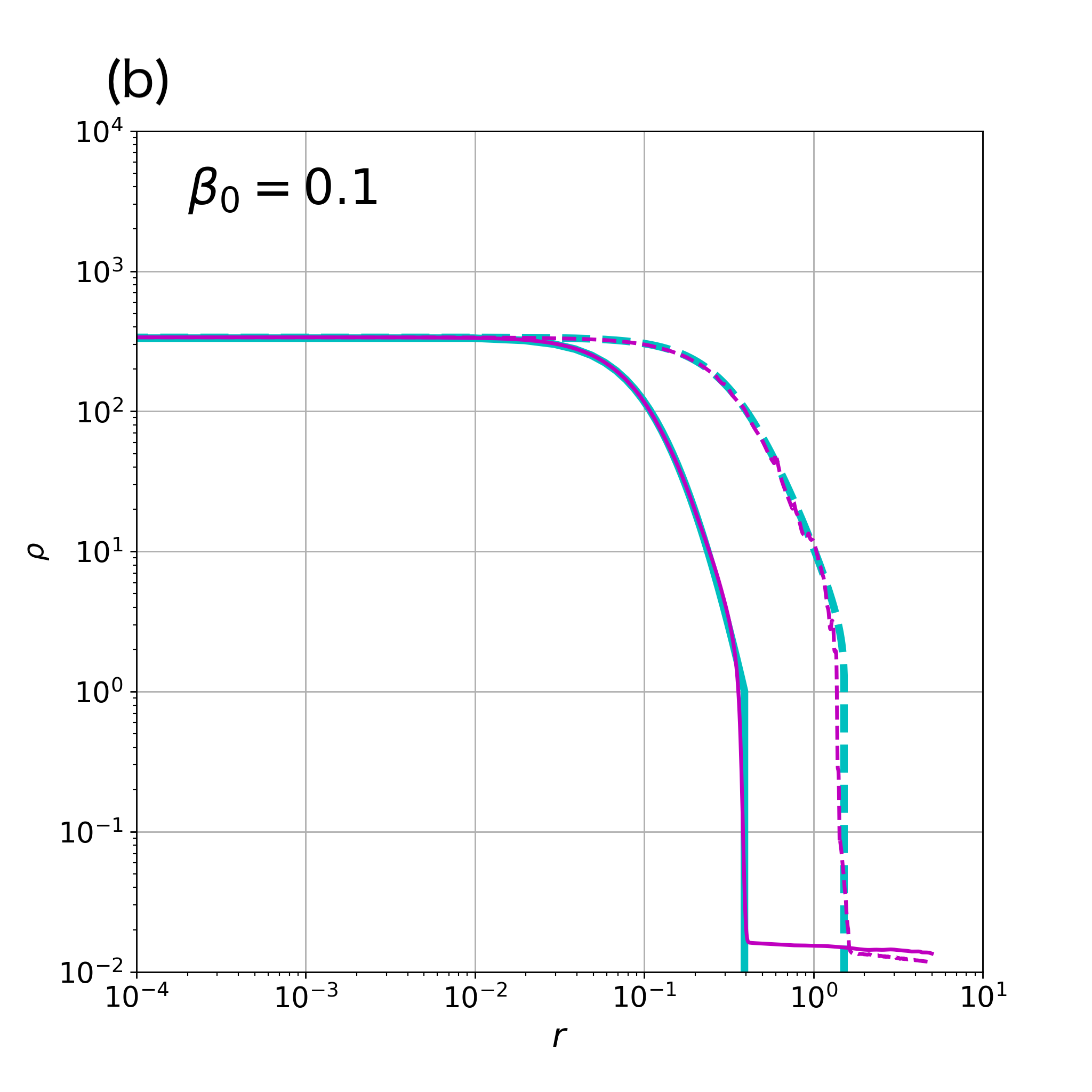}& \includegraphics[keepaspectratio,scale=0.3]{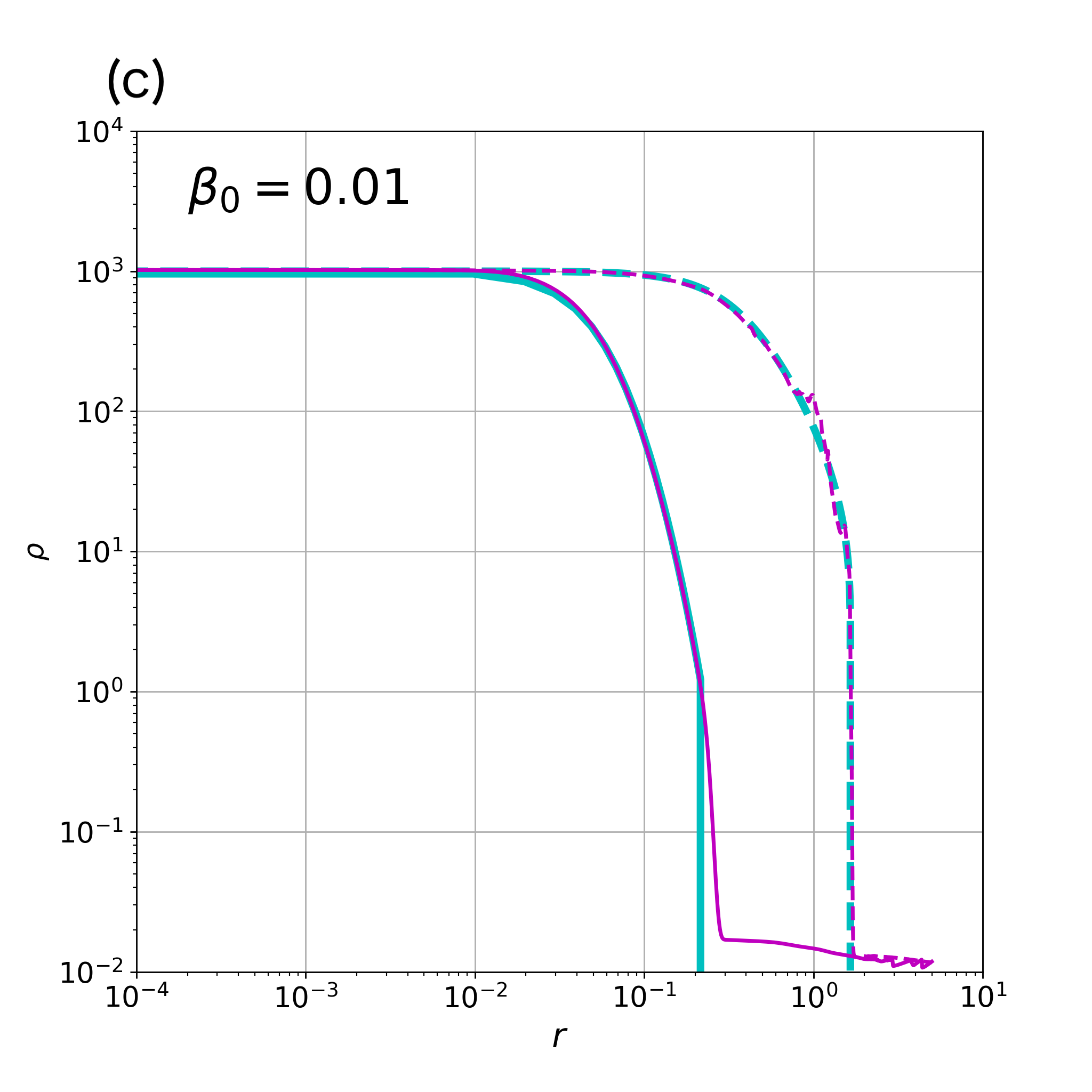}\\ 
     \end{tabular}
      \caption{ 
      Same as Figure \ref{fig:density_L05B01}, but for the comparison of models with different $\beta_0$ values.
      Panels (\textit{a}), (\textit{b}), and (\textit{c}) correspond to the models with $\beta_0=1$, $0.1$, and $0.01$, respectively.
      }
     \label{fig:L05_beta_density}
\end{figure*}

In the previous sections \ref{sec:collapse} and \ref{sec:stable}, we show the models with a plasma beta value of $\beta_0=0.1$.
Here, we compared the models with plasma beta values of $\beta_0=1$ (models $\beta 1$S and $\beta 1$M) and $0.01$ (models $\beta 001$S and $\beta 001$M).

Figure \ref{fig:beta_maxden} shows the maximum density evolution of the radial collapse (\textit{a}) and stable (\textit{b}) models.
For the radial collapse model (\textit{a}), the strength of the initial magnetic field does not affect the outcome of radial collapse. 
However, the stable model (\textit{b}) maintains a stable state while undergoing oscillations, regardless of the initial plasma beta value. 
We assumed that the line masses of intermediate (M) and less-massive (S) filament are 0.75 and 0.5 times as large as the magnetically critical line mass of Equation (\ref{eq:dimensionless_magnetized_critical_line_mass}), respectively. 
As the magnetically critical line mass increases with a decrease in plasma beta $\beta_0$, line masses of the respective models satisfy the following relations: $\lambda(\beta\mathrm{1M}) < \lambda(\beta\mathrm{01M}) < \lambda(\beta\mathrm{001M})$, and $\lambda(\beta\mathrm{1S}) < \lambda(\beta\mathrm{01S}) < \lambda(\beta\mathrm{001S})$.
The line mass increases as the plasma beta value decreases, leading to a stronger gravitational force in the merged filament. 
Consequently, the radial collapse model shows a shorter timescale, and the stable model exhibits a shorter oscillation period. 
We defined a time for the maximum density reaching $\rho_\mathrm{lim}$ as $t_\mathrm{lim}$. 
In the radial collapse model, $t_\mathrm{lim}$ changes from $t_\mathrm{lim}=1.103$($\beta1$M) to $t_\mathrm{lim}=0.443$($\beta001$M). 
In the stable model, the oscillation period $T_\mathrm{osi}$ becomes shorter from $T_\mathrm{osi}\simeq 4.0$($\beta1$S) to $T_\mathrm{osi}\simeq 1.8$($\beta001$S).

Next, we study the density structure of both models.
Figure \ref{fig:L075_beta_density} shows the comparison of the density profiles for the radial collapse models with varying plasma beta values ($\beta_0$).
The profiles are shown for the final epochs, which have the same central density of $\rho_c=\rho_\mathrm{lim}\simeq 1.66\times 10^6$. 
In Figure \ref{fig:L075_beta_density}, two analytic solutions are shown, that of the self-gravitating isothermal plane-parallel gas disk (solid line; $\rho=\rho_c \mathrm{sech}^2\left(r\sqrt{\rho_c/2}\right)$)  and that of the self-gravitating isothermal cylinder (dashed line; Equation (\ref{eq:iso_density})).
By considering the central high-density part $\rho\gtrsim 10^4$ in Figure \ref{fig:L075_beta_density}(\textit{a}), we realize that the density profiles on the $x$-axis are more similar to the hydrostatic solution of the gas disk (solid line) than the solution of the gas cylinder (dashed line), regardless of $\beta_0$.
This is because the Lorentz force is weak along the $x$-axis, leading to the thermal pressure 
playing a dominant role against the self-gravity of the gas disk.
On the contrary, in the central high-density part on the $y$-axis (\textit{b}), the profiles seem to be more expanded than that of the isothermal gas cylinder as the plasma beta value decreased from $\beta_0=1$ (red) to 0.01 (yellow) owing to the strong Lorentz force.

Figure \ref{fig:L05_beta_density} shows the density distribution of models $\beta1$S (\textit{a}), $\beta01$S (\textit{b}), and $\beta$001S (\textit{c}) achieved at the respective final epochs as $t=8.100$, 6.600, and 4.800. 
Figure \ref{fig:L05_beta_density} shows the comparison of these stable models with the magnetohydrostatic equilibrium states described in \citet{2014ApJ...785...24T}. 
Of note, the equilibrium state is selected to have the same magnetic flux and the same central density as the final state of the corresponding stable model. 
The density structures in all the models resemble the equilibrium states; in particular, in the models with a strong magnetic field (\textit{b}) and (\textit{c}), two profiles are well-matched. 
Thus, in general, less massive filaments lead to a merged filament stable with oscillation, and its density profile is highly similar to that of the equilibrium state.

\subsection{Effect of Initial Velocity}\label{sec:velocity}

\begin{figure*}
    \centering
     \begin{tabular}{cc}
         \includegraphics[keepaspectratio,scale=0.4]{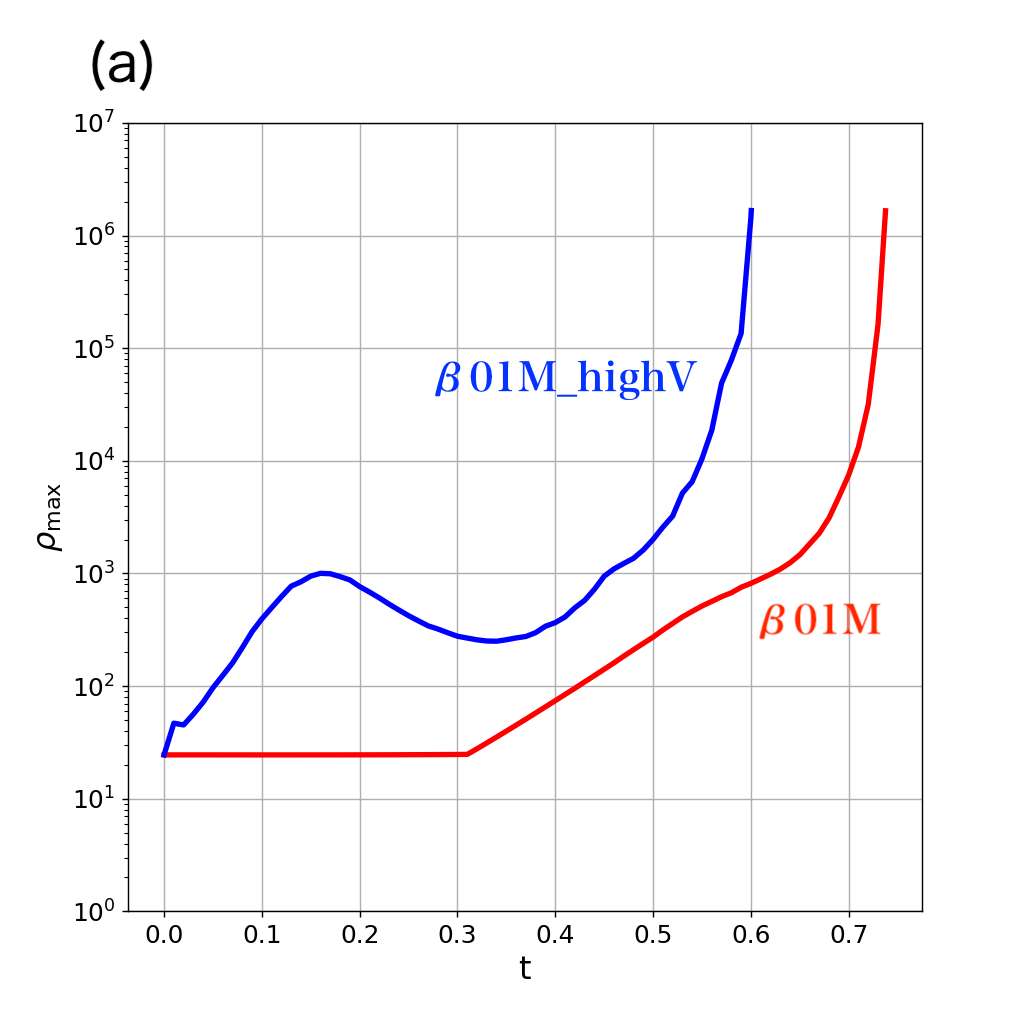} & \includegraphics[keepaspectratio,scale=0.4]{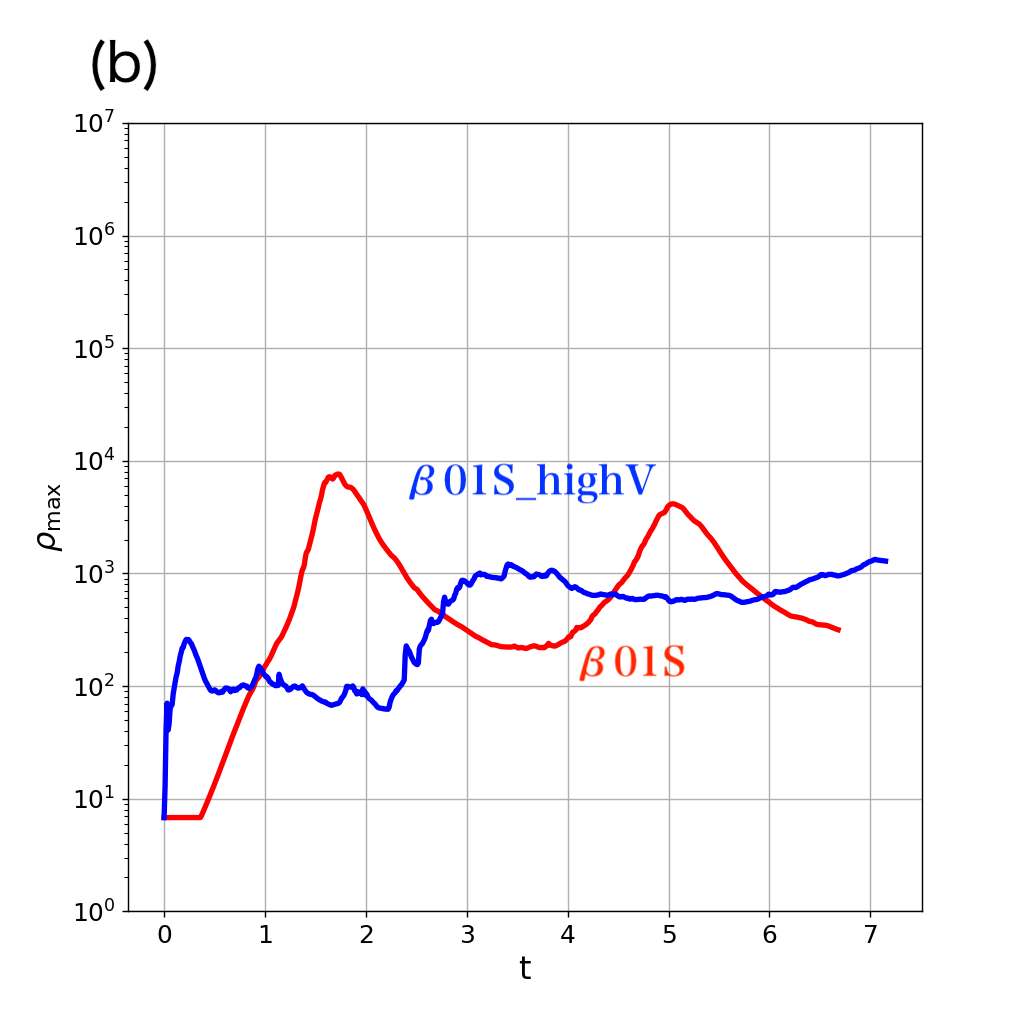}\\ 
     \end{tabular}
      \caption{
      Same as Figure \ref{fig:beta_maxden}, but for the comparison of models with different $V_{\rm int}$ values. 
      Maximum density is plotted against the time.
      Panels (\textit{a}) and (\textit{b}) show the results for $\beta$01M and $\beta$01M\_highV, and $\beta$01S and $\beta$01S\_highV, respectively.  
      Color represents the initial relative velocity as $V_{\rm int}=1$ (red) and $V_{\rm int}=10$ (blue). 
      The plasma beta value is set to $\beta_0=0.1$.
      }
     \label{fig:Vint_maxden}
\end{figure*}

\begin{figure*}
    \centering
     \begin{tabular}{cc}
         \includegraphics[keepaspectratio,scale=0.4]{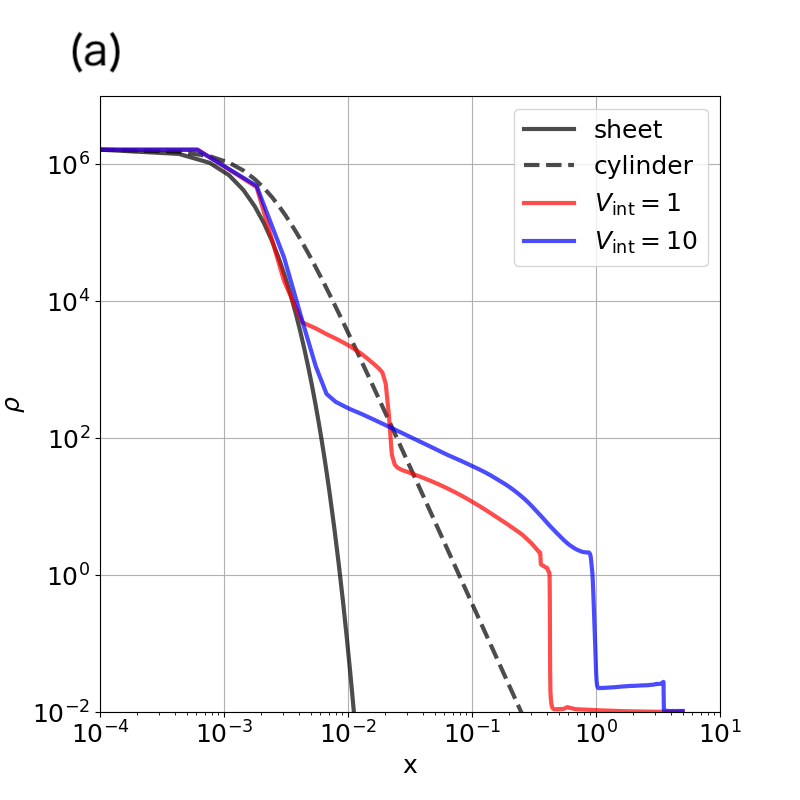} & \includegraphics[keepaspectratio,scale=0.4]{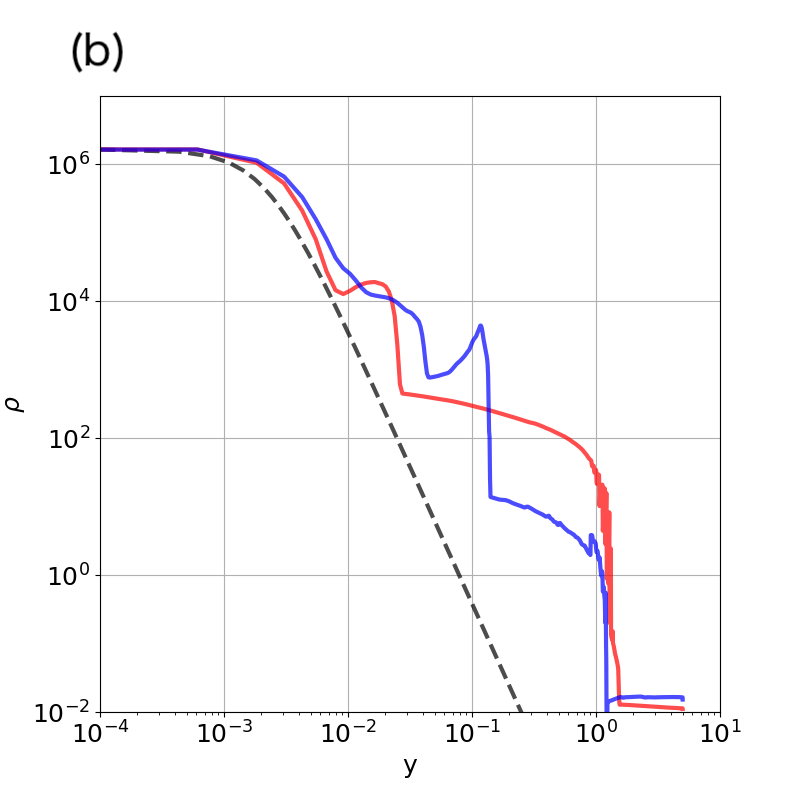}\\  
     \end{tabular}
      \caption{ 
      Same as Figure \ref{fig:L075_beta_density}, but for the comparison of models $\beta$01M and $\beta$01M\_highV with different initial relative velocities of $V_{\rm int}=1$ (red) and $V_{\rm int}=10$ (blue). 
      Other parameters are the same: $\lambda_\mathrm{tot}=2.87$ and $\beta_0=0.1$.
      Images are captured for the final epoch with $\rho_c=\rho_\mathrm{lim}=1.66\times 10^6$.}
     \label{fig:L075_Vint_density}
\end{figure*}

\begin{figure}
    \centering
     \begin{tabular}{c}
         \includegraphics[keepaspectratio,scale=0.4]{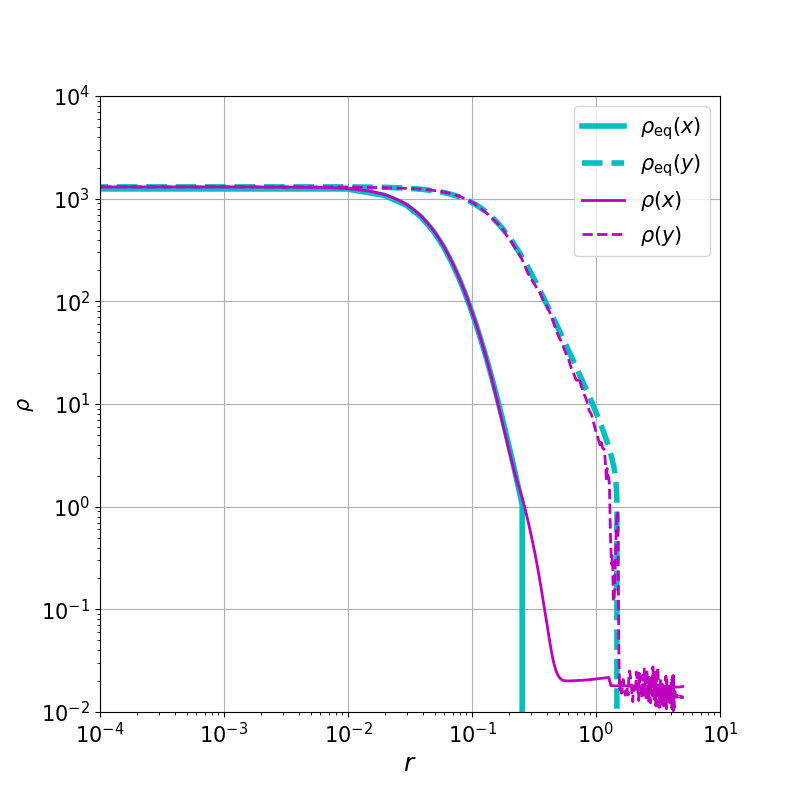}  
     \end{tabular}
      \caption{ 
      Same as Figure \ref{fig:density_L05B01}, but for a model with an initial relative velocity of $V_{\rm int}=10$ ($\beta$01S\_highV). 
      The density profile at $t=7.10$ is illustrated. 
      The other parameters are the same as the model $\beta$01S shown in Figure \ref{fig:density_L05B01}: $\lambda_\mathrm{tot}=1.91$ and $\beta_0=0.1$.
      }
     \label{fig:L05_Vint_density}
\end{figure}

In this section, we describe the evaluation of the effect of the initial velocity on the evolution of filament collision. 
We compare four models: $\beta$01M, $\beta$01M\_highV, $\beta$01S, and $\beta$01S\_highV. 
In model $\beta$01M\_highV, we assume the same filament (line mass and the magnetic flux) as model $\beta$01M but with a larger collision velocity, $V_\mathrm{int}=10$.
Furthermore, model $\beta$01S\_highV is a high-speed collision model of $\beta$01S.
Figure \ref{fig:Vint_maxden} shows the evolution of the maximum density for these four models.
Models $\beta$01M\_highV and $\beta$01M exhibit runaway collapse, and models $\beta$01S\_highV and $\beta$01S result in oscillation in $\rho_\mathrm{max}$.

Panel (\textit{a}) shows the effect of the initial velocity on the time scale of reaching the radial collapse. 
We defined a time for the maximum density reaching $\rho_\mathrm{lim}$ as $t_\mathrm{lim}$.
The time is approximately equal to $t_\mathrm{lim}=0.735$ ($V_\mathrm{int}=1$) and $0.600$ ($V_\mathrm{int}=10$), respectively. 
Thus, we show that the time scale required to reach radial collapse $t_\mathrm{lim}$ decreases as the initial velocity increases. 
This is because a higher initial velocity leads to a more efficient inflow of gas toward the shocked region. 
In other words, the time required for the shock wave to sweep the filament becomes shorter compared with a low-speed collision.
However, if the initial velocity is too large (e.g., $V_\mathrm{int}\gg10$), it is likely for the merged filament to expand more before undergoing radial collapse due to its momentum.  
We confirmed this finding using a model colliding with an extreme collision speed $V_\mathrm{int}=15$ (not shown in the paper).  
Such high-speed collisions result in a long time to collapse compared with the case with $V_\mathrm{int}=1$.
By contrast, panel (\textit{b}) shows that, even by considering the supersonic collision with a high Mach number ($V_\mathrm{int}=10$), less massive filaments of $\beta01$S\_highV model evolve into a stable state and form a static filament.

Next, we focus on the density structure of the radial collapse and stable models with a high Mach number ($V_\mathrm{int}=10$). 
Figure \ref{fig:L075_Vint_density} shows the density profile on the $x$- and $y$-axes of the intermediate-mass model, $\beta01$M\_highV. 
By comparing the models with $V_{\rm int}=1$ (red line) and that with $V_{\rm int}=10$ (blue line), we observed that, despite the high initial velocity, the filament width of the shocked region remains similar to that of the result for $V_{\rm int}=1$. 
In other words, the structure of the dense part ($\rho \gtrsim 10^4$) of the shocked region remains almost similar, regardless of the initial velocity.

In addition, Figure \ref{fig:L05_Vint_density} shows the density profile on the $x$- and $y$-axes of the less-massive model ($\beta01$S\_highV). 
The structure of the shocked region overlaps with the magnetohydrostatic equilibrium state, similar to that of $V_{\rm int}=1$ (see Figure \ref{fig:density_L05B01}).

\section{Discussion}\label{sec:discussion}
\begin{figure}
    \centering
    \includegraphics[keepaspectratio,scale=0.35]{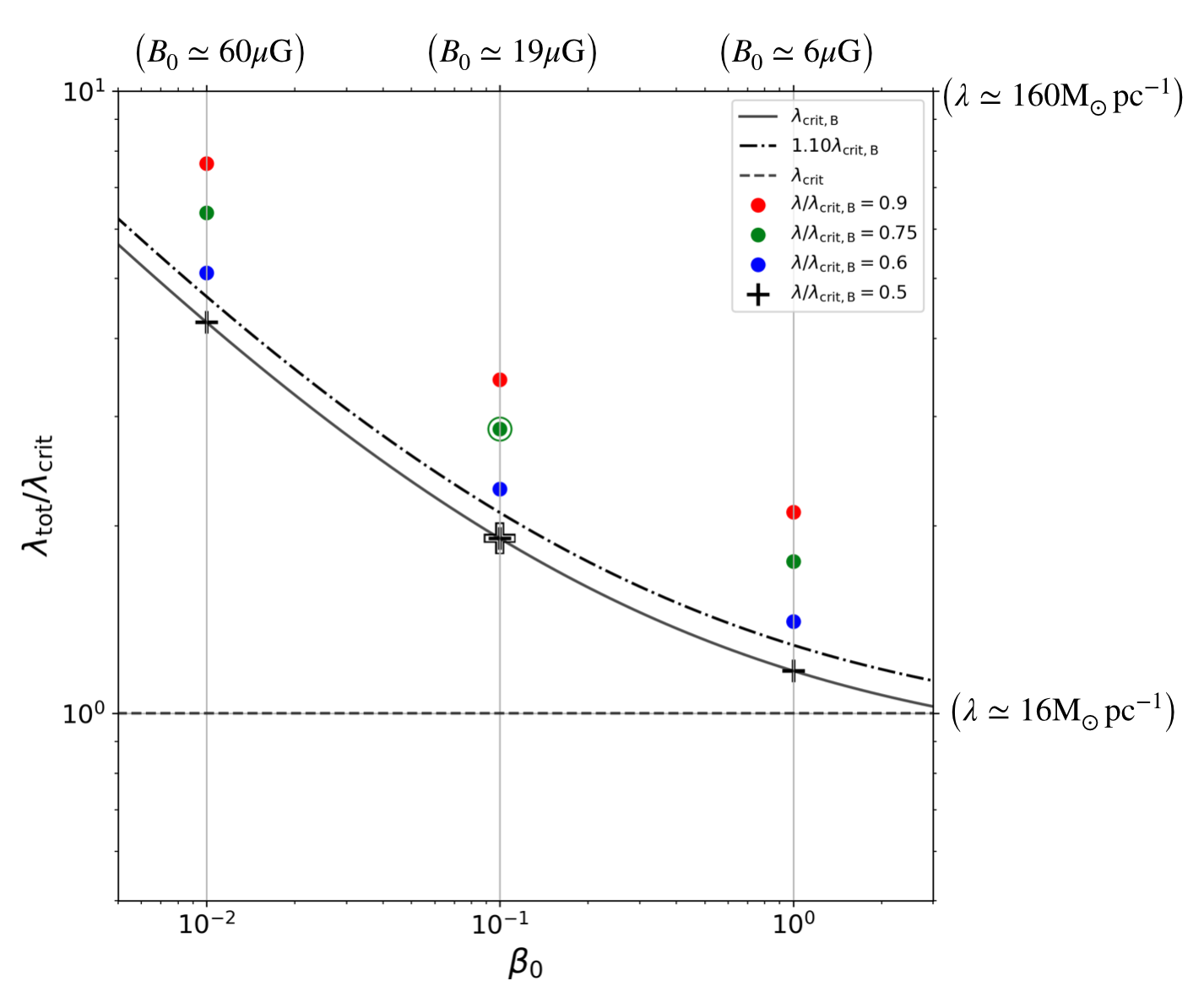}
      \caption{Criteria for the radial collapse
      displayed on the plane ($\lambda_{\rm tot},\beta_0$).
      The circles and cross points represent the radial collapse and stable models, respectively.
      The enclosed symbols correspond to models with large initial velocities ($V_\mathrm{int}=10$).
      Color represents different line masses normalized by $\lambda_\mathrm{crit,B}$: $\lambda/\lambda_\mathrm{crit,B}=0.9$ (red), 0.75 (green), 0.6 (blue), and 0.5 (black).
      The solid line corresponds to the magnetized critical line mass \citep{2014ApJ...785...24T}.
      The dashed horizontal line represents the critical line mass of the nonmagnetized filament \citep{1963AcA....13...30S,1964ApJ...140.1056O}.
      The dashed line represents the critical line mass obtained by considering the geometric mean of the lightest line mass in the radial collapse and stable models.
       We also show the dimensional quantities for $\lambda_\mathrm{tot}[\mathrm{M_\odot{pc}^{-1}}]$(right) and $B_0[\mathrm{\mu G}]$(upper), where the sound speed is $190\mathrm{m\,s^{-1}}$ at $T=10 \mathrm{K}$ and the filament surface number density is $n_s=10^3 \mathrm{cm^{-3}}$.
      }
    \label{fig:lambda_vs_beta}
\end{figure} 

\subsection{Criteria for Radial Collapse}
\label{subsection:criteria}

In this study, we attempt to determine the conditions under which the merged filament becomes unstable in the radial direction by filament collision. 
Figure \ref{fig:lambda_vs_beta} summarizes the results of the collision, indicating whether the merged filament underwent radial collapse or not.
This shows that models with larger line masses and/or models with a weaker magnetic field (larger $\beta_0$) are likely to experience radial collapse.
The outcome does not appear to depend on the collision velocity, $V_\mathrm{int}$. 
In Figure \ref{fig:lambda_vs_beta}, we plotted the critical line mass of magnetized filaments, $\lambda_\mathrm{tot}=\lambda_\mathrm{crit,B}(\beta_0)$.
The results can be separated by comparison with the magnetically critical line mass (Equation (\ref{eq:dimensionless_magnetized_critical_line_mass})).
Merged filaments with $\lambda_\mathrm{tot} \gtrsim\lambda_\mathrm{crit,B}$ undergo radial collapse, whereas those with $\lambda_\mathrm{tot} \lesssim\lambda_\mathrm{crit,B}$
exhibit stable oscillation.
From this finding, we conclude that the necessary condition for the occurrence of radial collapse in the filament collision is that the total line mass must exceed the magnetically supported critical line mass.
In other words, the merged filament should be a magnetically supercritical filament.
Even if the magnetic field is extremely strong, the magnetically supercritical filament is likely to undergo radial collapse. 
If the collision velocity is too fast, as explained in Section \ref{sec:velocity}, collapse may be delayed. 
Therefore, the collapse appears not to occur within a finite time for the collision with an ultimately large initial velocity. 
However, in the case of a filament collision originated from typical velocity dispersion ($\Delta v \lesssim \mathrm{a\,few\, km\,s^{-1}}$) in molecular clouds, as considering in this study, this delayed collapse is not realized.

It is evident that the critical line mass is a dominant factor in determining the radial instability of the shocked region.
This conclusion may be restricted to a specific scenario of a head-on collision described here.
In such a collision, a merged filamentary structure elongated in the $z$ direction is retained as the shocked region because there is no inclination between the long axes of two filaments.
As a result, we expect the radial instability of the shocked region to be described by the magnetically critical line mass (Equation (\ref{eq:magnetized_critical_line_mass})).
This finding can help in considering the evolution of a filament--filament collision in a more general configuration.
An elongated merger of a filament--filament collision is made even by a collision with some inclinations, although the filaments far from the collision point pass by freely. 
When such colliding filaments share the same magnetic flux, the necessary condition of the gravitational collapse given as the merged filament (or a hub) is magnetically supercritical.

\subsubsection{Comparison with Collision between Spherical Clouds}

While our calculations did not show a velocity dependence, it has been reported that in the simulation studies which investigate collisions between spherical clouds \citep{1984ApJ...279..335G,1987PThPh..78.1250N}.
They found that the critical total mass above which gravitational collapse is triggered depends strongly on the collision velocity. 
When the collision velocities are low, the shocked region undergoes collapse if the total mass exceeds the Bonnor-Ebert mass \citep{1955ZA.....37..217E,1956MNRAS.116..351B}, which is the critical mass of the isothermal equilibrium cloud.
On the other hand, if clouds collide with a high velocity, the shocked region expands and does not show collapse even when the mass exceeds the Bonnor-Ebert mass.
Therefore, head-on collisions of spherical clouds which exceed the Bonnor-Ebert mass will collapse if the collision velocity is low, but will not collapse if the collision velocity is too high. 
We suspect that this difference is due to the distance dependence of gravity for spherical ($\propto r^{-2}$) and cylindrical ($\propto r^{-1}$) coordinates. Thus, in the spherical case, the gas dispersed by the collision easily escapes the gravity of the shocked region compared to the cylindrical case. 

In addition to this, although, if the collision velocity is too large, the shocked region will turn to expansion, the shocked region shows collapse even with the total mass below the Bonner-Ebert mass when the collision velocity is moderate. 
This fact in the moderate collision velocity regime comes from that the critical mass of the sphere depends on the external pressure ($\propto p_\mathrm{ext}^{-1/2}$), and thus the critical mass of the shocked region decreases due to the increasing external pressure, which is derived from the collision velocity.
On the other hand, the critical line mass of a cylinder is independent of the external pressure. 
Consequently, the collision velocity is less important to the stability of the shocked region in contrast to the spherical configurations.

Although the filaments have a strong gravity compared to the spheres, which makes it difficult for the dispersed gases to escape, if we had considered collisions at higher velocities than those considered in our paper, we might have seen the velocity dependence of the threshold line mass. However, if the origin of the collision velocity is inherited from the turbulence of the molecular clouds,  then the collision speed we assumed ($V_\mathrm{int}/c_s=1-10$) is reasonable.

\subsubsection{Comparison with Observations}
Aquila Serpens south region is one of the typical examples where stars are formed in the filament (section \ref{sec:intro}). 
This region contains three clumps (filaments) \citep{2013ApJ...778...34T}, and the main filament is regarded as a site where multiple filaments collide with each other and star formation is in progress in the merged filament \citep[Figure 3 of][]{2014ApJ...791L..23N}.
From the gas mass of the main filament $M\sim 230 M_\odot$, the line mass is estimated as $\lambda_\mathrm{POS}\sim 480 M_\odot \mathrm{pc}^{-1}$ by assuming a length of 0.48 pc 
\footnote{
Figure 2 of \citet{2013ApJ...778...34T} shows the length of the main filament and the northern clump to be approximately 4 arcminutes. We adopted a distance of $415 \mathrm{pc}$ to Serpens South, resulting in a length of $0.48 \mathrm{pc}$.
}
projected on the plane of the sky (POS).
The northern clump has a similar mass ($M\simeq 190 M_\odot$) and size ($0.48\mathrm{pc}$) to the main clump but are composed of two filaments, which appear to collide in the future.

Magnetic field strength is measured using the Davis--Chandrasekhar-Fermi method \citep{1951PhRv...81..890D,1953ApJ...118..113C}.
\citet{2020NatAs...4.1195P} reported an estimated magnetic field strength of approximately $870\mu \rm G$ within the main filament of the region based on far-infrared observations. 
On the contrary, \citet{2019PASJ...71S...5K} estimated the magnetic field strength in the same region to be ranging from $10\mu \mathrm{G}$ to $80\mu \mathrm{G}$ using near-infrared wavelengths, which trace the magnetic field strength in a relatively low-density medium.
These observations are consistent with our simulation results, in which the magnetic field strength is amplified by approximately two orders of magnitude due to the collision compared with the initial filament.

Thus, in the Aquila Serpens south region, two filaments with $\lambda_\mathrm{POS}\sim 200 M_\odot \mathrm{pc}^{-1}$ collide with each other and form a merged filament with $\lambda_\mathrm{POS}\sim 400 M_\odot \mathrm{pc}^{-1}$.
Although there is some uncertainty regarding the magnetic field strength, the magnetic field strength outside the main filament is estimated to be $B_\mathrm{POS}\sim 80~\mu \mathrm{G}$, which corresponds to $\beta_\mathrm{POS}=5.4\times 10^{-3}$ for an external pressure of $p_\mathrm{ext}=10^4\mathrm{K\,cm}^{-3}$.

As the magnetically critical mass is given as $\lambda_\mathrm{crit}(B_\mathrm{POS})\simeq 88 M_\odot\mathrm{pc}^{-1}$ from equation (\ref{eq:magnetized_critical_line_mass}), both the main and north filaments appear to be magnetically supercritical as $\lambda_\mathrm{POS} > \lambda_\mathrm{crit}(B_\mathrm{POS})$, where we assumed $R_0=0.2\mathrm{pc}$, $c_s=190\mathrm{m\,s}^{-1}$, and $n_s=10^3\mathrm{cm}^{-3}$.
However, we must also consider the geometrical projection effect.
setting the angle between the line-of-sight and the filament axis, $\alpha$,
 the true line mass and magnetic field strength are expressed with those of the POS as $\lambda=\lambda_\mathrm{POS}\sin{\alpha}$ and $B_0=B_\mathrm{POS}/\cos{\alpha}$ (or $\beta_0=\beta_\mathrm{POS}\cos^2\alpha$).
This reduces to $\lambda/\lambda_\mathrm{crit}(B_0)\simeq \sin\alpha\cos\alpha \left[\lambda_\mathrm{POS}/\lambda_\mathrm{crit}(B_\mathrm{POS})\right]$, when the second term is ignored in equation (\ref{eq:magnetized_critical_line_mass}).
As the average of $\langle\sin\alpha\cos\alpha\rangle$ is 1/3, the correction factor is expected to be approximately 1/3.
After considering the geometrical correction, although the main filament remains slightly supercritical, the two north filaments become subcritical.
In conclusion, the condition for radial collapse is satisfied in the collision, which we observed in the Aquila Serpens south region.

\subsection{Self-similarity of the Collapsing Model}\label{sec:self-similar}
\begin{figure*}
    \centering
     \begin{tabular}{cc}
         \includegraphics[keepaspectratio,scale=0.5]{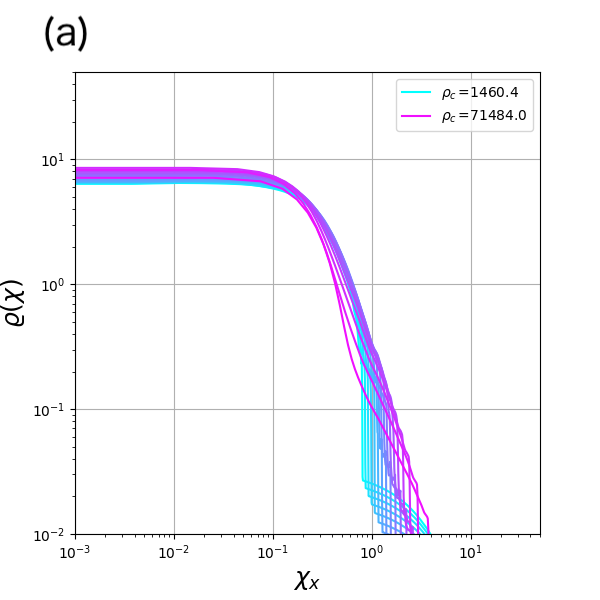} & \includegraphics[keepaspectratio,scale=0.5]{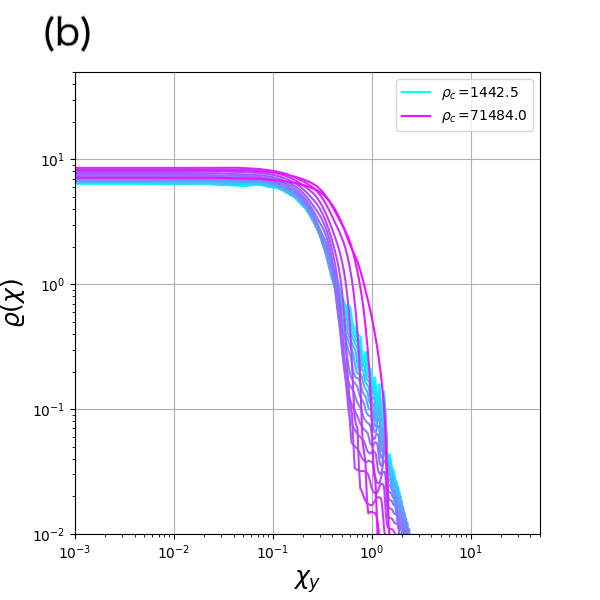}\\  
     \end{tabular}
      \caption{ 
      Density profiles on the $x$- and $y$-axes for the radial collapse models ($\beta1$M), in which $\beta_0=1$ and $V_{\rm int}=1$.
      The ordinate denotes $\varrho$ and the abscissa denotes $x$- and $y$-components of $\bm{\chi}$.
      }
     \label{fig:Self-similarity}
\end{figure*} 
In this section, we describe the examination of self-similar solutions of a collapsing filament.
This solution has been previously studied to describe the properties of the collapsing filament in \citep{1998PASJ...50..577K}.
In \citet{1998PASJ...50..577K}, for similarity coordinates, the defined zooming coordinate $\bm{\chi}$ and the density in the zooming coordinate $\varrho$ as,
\begin{equation}
   \bm{\chi}\equiv \frac{\bm{r}}{c_s|t-t_0|},
\end{equation}
\begin{equation}
   \varrho(\chi)\equiv 2\pi G(t-t_0)^2\rho,
\end{equation}
where $t_0$ represents the time when the central density becomes infinity.\footnote{
To determine $t_0$, we first examine the time evolution of $1/\sqrt{\rho_\mathrm{max}}$.
We identify the part that follows a linear relation similar to$1/\sqrt{\rho_\mathrm{max}}=at+b$.
The time at which $1/\sqrt{\rho_\mathrm{max}}=0$ corresponds to $t_0=-b/a$.
}

Figure \ref{fig:Self-similarity} shows the time evolution of the density profiles written in the similarity coordinate $(\bm{\chi},\varrho(\bm{\bm{\chi}}))$ for model $\beta$1M.
The color gradation indicates a time sequence as the central density of the merged filament increases from $\rho_c=1460$ (cyan) to $\rho_c=71484$ (magenta).
Panel (\textit{a}) shows the density profiles on the $x$-axis.
Although $\rho_c$ varies by a factor of $\sim 50$, 
the density profiles appear not to depend on time significantly when the similarity coordinates are used.
Panel (\textit{b}) shows the density profiles on the $y$-axis.
Compared with the density profiles on the $x$-axis, the density profiles on the $y$-axis demonstrated an increase in their width as the central density increased.
Thus, although the convergence to the self-similarity depends on the direction, we expect that the contraction proceeds in a self-similar manner, especially for the collision direction.

\subsection{Initial Line Mass Dependence of the Shocked Region}
\begin{figure}
    \centering
     \begin{tabular}{c}
         \includegraphics[keepaspectratio,scale=0.4]{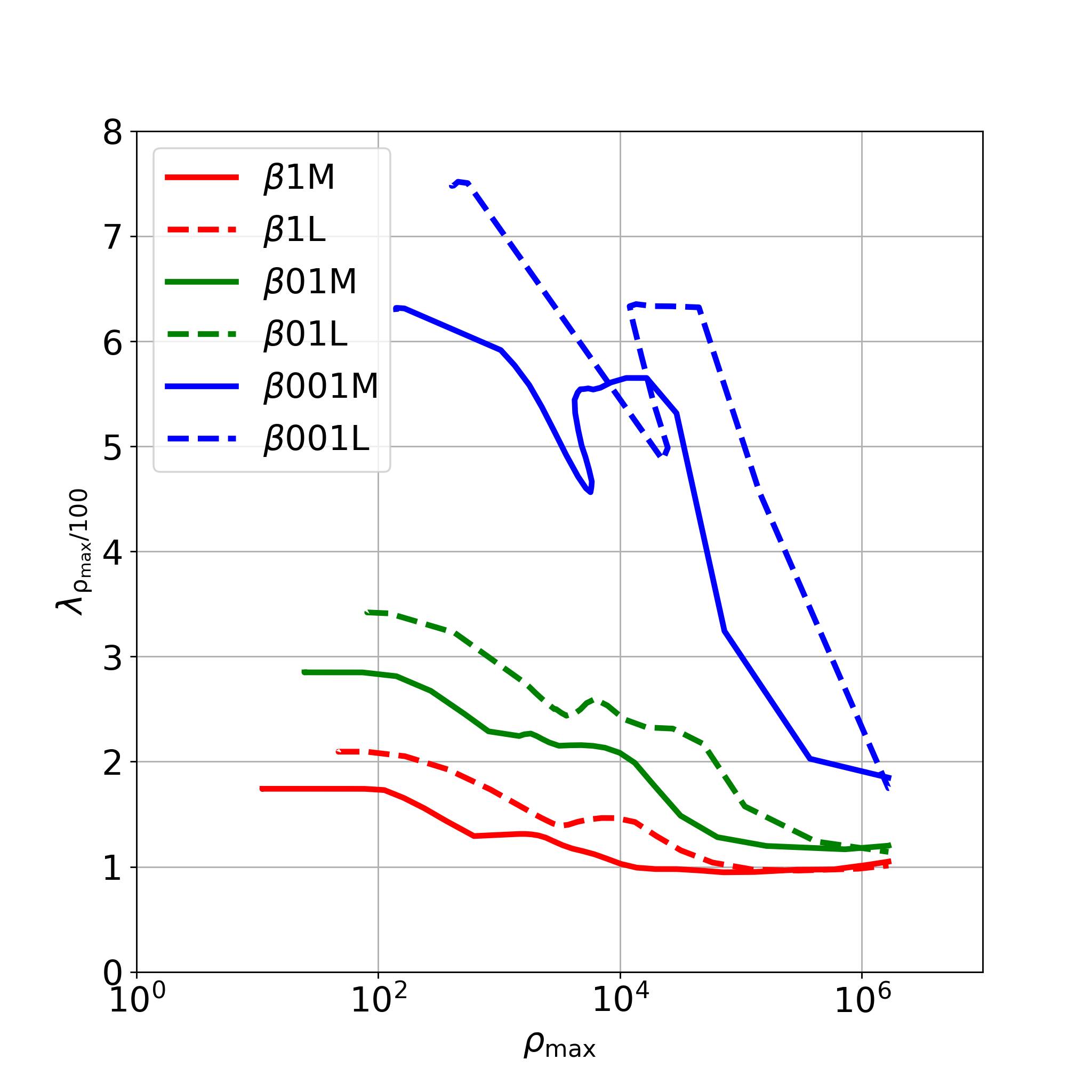}
     \end{tabular}
      \caption{
      Line mass at the shocked region plotted against the maximum density.
      The $y$-axis is $\lambda_{\rho_\mathrm{max}/100}$, which is the line mass that is counted over $\rho_\mathrm{max}/100$.
      The $x$-axis shows the maximum density.
      The solid and dashed lines represent the intermediate and massive line mass models, respectively, and the different colors indicate models with $\beta_0=1$ (red), 0.1 (green), and 0.01 (blue), respectively.
      The initial relative velocity is fixed as $V_{\rm int}=1$.
      }
     \label{fig:initial-linemass}
\end{figure}

In this section, we describe the examination of the initial line mass dependence of the shocked region.
Here, we assume that the line mass of the shocked region $\lambda_{\rho_\mathrm{max}/100}$ is estimated by integrating the density larger than $\rho_c/100$ as $\lambda_{\rho_\mathrm{max}/100}\equiv\iint_{\rho\ge\rho_\mathrm{max}/100 }\rho dx dy$.
This assumption aims to follow the evolution of the dense part in the shocked region. 
We compare the models with the same plasma beta value $\beta_0$ but different initial line masses $\lambda_\mathrm{tot}$ for three pairs of models: ($\beta1$M, $\beta1$L), ($\beta01$M, $\beta01$L), and ($\beta001$M, $\beta001$L).

In Figure \ref{fig:initial-linemass}, the line mass of the shocked region $\lambda_{\rho_\mathrm{max}/100}$ is plotted against the maximum density $\rho_\mathrm{max}$.
As the maximum density rarely decreases with time, $\rho_\mathrm{max}$ substitutes for the elapsed time $t$.
The initial relative velocity is set to $V_{\rm int}=1$.
Although the line mass of the dense part $\lambda_{\rho_\mathrm{max}/100}$ differs initially (left end of each line), both line masses $\lambda_{\rho_c/100}$ come close to almost the same value during the time evolution for all pairs of different plasma beta values (right end of each line).
In particular, in the models with $\beta_0=1$ (red line), the line mass maintains a constant value $\lambda_{\rho_\mathrm{max}/100}\simeq \lambda_\mathrm{crit}$ after the maximum density exceeds $\rho_\mathrm{max}\gtrsim 10^5$, regardless of the initial line mass.

In the model with a small plasma beta value of $\beta_0=0.01$, the final $\lambda_{\rho_\mathrm{max}/100}$ of the two models agrees with each other.
However, it is unclear whether $\lambda_{\rho_\mathrm{max}/100}$ maintains the final value for further contraction.
Evolution may reach only the extremely early stage with small beta models; thus, we require further evaluation to confirm the true convergence in the models with $\beta_0=0.01$.   
In conclusion, although the degree of convergence depends on the magnetic field strength, the typical line mass of the high-density portion of the contracting filament $\lambda_{\rho_\mathrm{max}/100}$ does not depend on the initial line mass of the filament.

\begin{figure*}
    \centering
     \begin{tabular}{cc}
         \includegraphics[keepaspectratio,scale=0.4]{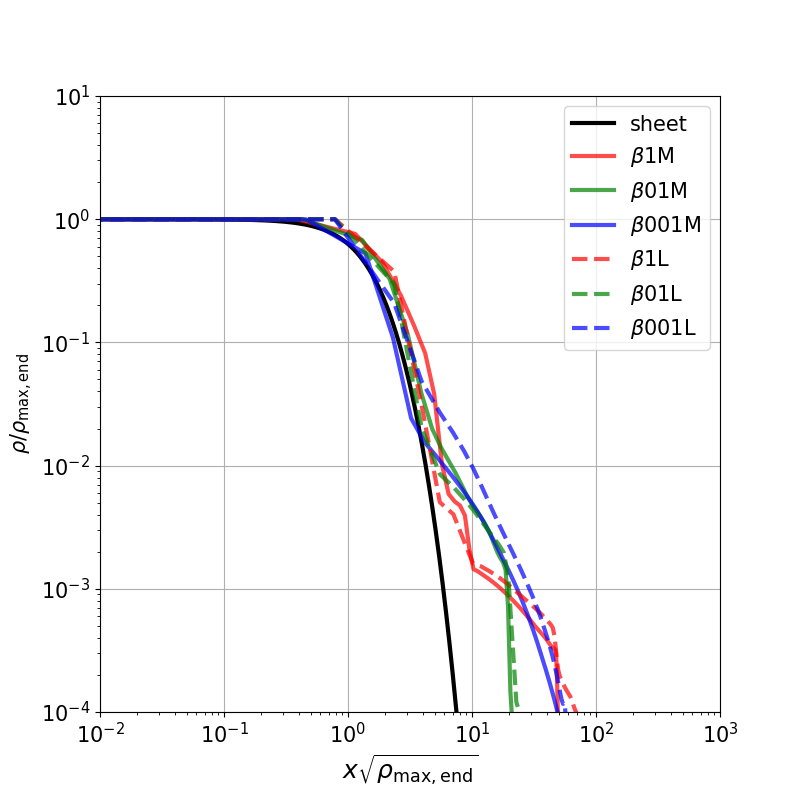}& \includegraphics[keepaspectratio,scale=0.4]{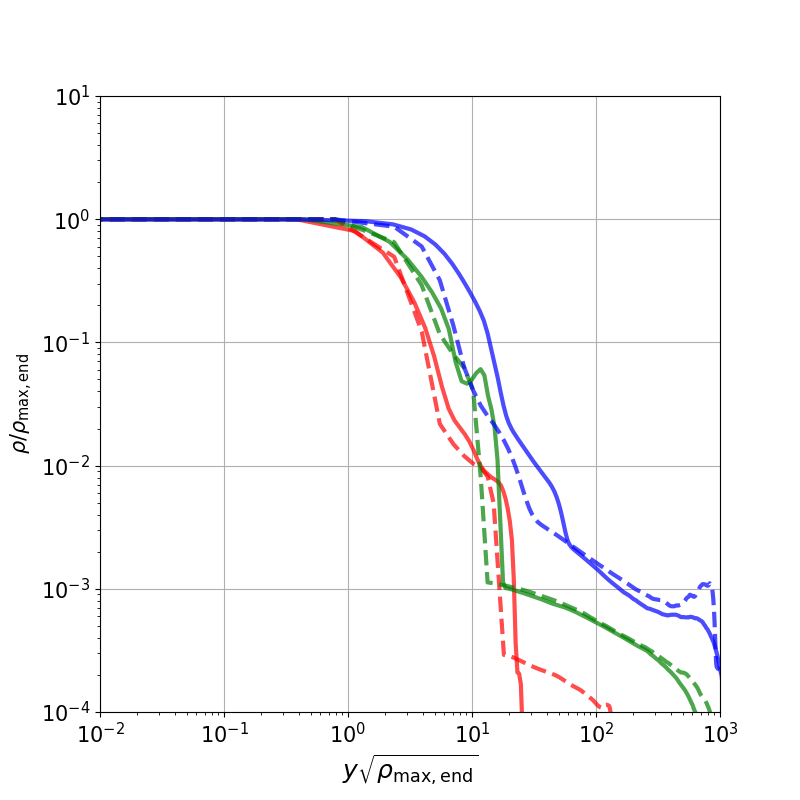}
     \end{tabular}
      \caption{ 
       Comparison of the density profiles on the $x$- and $y$-axes for the radial collapse models with different initial line masses. A comparison of the three pairs of models, ($\beta$1M, $\beta$1L), ($\beta$01M, $\beta$01L), and ($\beta$001M, $\beta$001L), is performed.
      The vertical axis represents the relative density to the maximum density at the final epoch, $\rho/\rho_\mathrm{max,end}$.
      The horizontal axis represents the distance normalized by the scale-height at the center of the filament, $\tilde{L}=c_s/(4\pi G \rho_\mathrm{max,end})^{1/2}$.
      Color represents the plasma beta value: red ($\beta_0=1$), green ($\beta_0=0.1$), and blue ($\beta_0=0.01$). 
      The solid and dashed lines represent the models with medium (M) and heavy (L) line masses. 
      The black line in the left panel corresponds to the hydrostatic density profile of an isothermal sheet.
      }
     \label{fig:initial-linemass-density}
\end{figure*}

Figure \ref{fig:initial-linemass-density} shows a comparison of the density profiles realized in the final state.
The density distribution is shown to be normalized by the maximum density at the final state such that the vertical axis represents $\rho/\rho_\mathrm{max}$.
In the horizontal axis, we consider the distance to be normalized by the scale-length given at the center as $x'=x/\tilde{L}$, where $\tilde{L}=c_s/(4\pi G \rho_\mathrm{max})^{1/2}$.
Figure \ref{fig:initial-linemass-density}(\textit{a}) shows that the dense part of $\rho/\rho_\mathrm{max}\gtrsim 10^{-2}$ has an extremely similar density distribution along the $x$-axis, which is shown in Figures  \ref{fig:L075_beta_density}(\textit{a}) and \ref{fig:L075_Vint_density}(\textit{a}).
By contrast, density distribution in the $y$-direction has a relatively poor similarity.
Models $\beta$1M and $\beta$1L have similar distributions in the $y$-direction.
This is valid for models $\beta$01M and $\beta$01L.
However, model $\beta$001M has a wider distribution than $\beta$001L. 
This implies that the time evolution is not sufficient for models with $\beta_0=0.01$ compared with the other two models with $\beta_0=1$ and 0.1 (Figure \ref{fig:initial-linemass}).

The convergence of $\lambda_{\rho_\mathrm{max}/100}$ (Figure \ref{fig:initial-linemass}) and the density distributions resembling each other (Figure \ref{fig:initial-linemass-density}) indicate that the collapse is expressed using the same solution, independent of the initial line mass after $\rho_c$ increases sufficiently.
Figure \ref{fig:initial-linemass} shows that the difference in $\lambda_{\rho_\mathrm{max}/100}$ owing to different $\beta_0$ also decreases as the collapse proceeds.
However, it is not clear whether the collapse is also expressed with a unique solution, irrespective of $\beta_0$, similar to the axisymmetric three-dimensional cloud \citep{1999ApJ...510..274N}.

\subsection{
Fragmentation of the merged filament.
}

In this section, we describe the investigation of the fragmentation of merged filaments resulting from head-on collisions.
Although our investigation primarily focused on the radial instability caused by such collisions, the fragmentation process is essential for the formation of molecular cloud cores from the filaments.
Studies on the fragmentation of magnetized filaments have been conducted by \citet{1963AcA....13...30S,1987PThPh..77..635N,1993ApJ...404L..83H,1993PASJ...45..551N,2000MNRAS.311..105F}, with the assumption of a magnetic field that is either parallel or helical to the filament axis. 
By contrast, \citet{2017ApJ...848....2H,2019ApJ...881...97H} assumed a magnetic field that is perpendicular to the filament axis.

For the nonmagnetized filament, the maximum growth rate of the gravitational instability was found to be $|\omega_\mathrm{max}|=0.339(4\pi G\rho_c)^{1/2}$ by \citet{1987PThPh..77..635N}, resulting in a timescale of the maximum growth as $t_\mathrm{min}\equiv 1/|\omega_\mathrm{max}|=2.94(4\pi G\rho_c)^{-1/2}$. 
If the dynamical timescale is slower than the timescale of the gravitational instability and if this continues for a sufficient duration, fragmentation occurs in the filament.
For a filament penetrated with a uniform magnetic field perpendicular to the axis, 
\footnote{Notably, although the model of a uniform magnetic field does not influence the cloud’s radial stability, it is not in the magnetohydrostatic equilibrium state.}
the wave number ($k$) and growth time showing the maximum growth are expressed approximately as $k\simeq0.15\sqrt{4\pi G\rho_c}/c_s$, and $t_\mathrm{min}\simeq 5.78(4\pi G\rho_c)^{-1/2}$ according to \citet{2017ApJ...848....2H}.
As the Lorentz force suppresses the fragmentation process, the growth rate becomes small, and the wavelength of the fragmentation becomes longer compared with the nonmagnetized ones.

We assume that density is perturbed along the $z$-direction and investigate whether it can grow and form fragmentations in the merged filament. 
For the radial collapse models, the central (maximum) density evolves almost monotonically, making fragmentation difficult (see Figures \ref{fig:maxden_L075_and_L05} and \ref{fig:beta_maxden}).
However, we find that the contraction stops temporally for a certain period, even in the radial collapse models such as $\beta$001M in Figure \ref{fig:beta_maxden}(\textit{a}) and $\beta$01M\_highV in Figure \ref{fig:Vint_maxden}(\textit{a}).
Here, we studied the possibility of whether gravitational instability developing during contraction is delayed.
For instance, model $\beta01$M\_{high}V expands after the collision at approximately $t=0.3$ (see Figure \ref{fig:Vint_maxden}). 
We found that the central density is maintained at $\rho'_c\simeq 250$ during $t'_\mathrm{stall}\simeq 0.08$, and the growth time of gravitational instability $t'_\mathrm{min}\simeq 0.36(\rho'_c/250)^{-1/2}$ is longer than the lag time of contraction ($t_\mathrm{stall}<t_\mathrm{min}$), and, even in the phase of maxima at $t\sim 0.16$, $t_\mathrm{stall}$ is shorter than $t_\mathrm{min}$. 
Another radial collapse model ($\beta001M$) with a bump is also $t_\mathrm{stall}<t_\mathrm{min}$ for the entire range of the simulation.

contrastingly, the merged filaments in the stable models can undergo fragmentation around the maxima or minima during oscillation, regardless of the plasma beta value.
For example, in the model of $\beta01$S, the maximum density is maintained at $\rho'_c\simeq 226$ for $t'_\mathrm{stall}\simeq 0.7$ (Figure \ref{fig:beta_maxden}(\textit{b})); thus, $t'_\mathrm{stall}> t'_\mathrm{min}\simeq 0.38(\rho'_c/226)^{-1/2}$.

\subsection{Effects of Different Geometries of the Magnetic Field Lines}

\begin{figure*}
    \centering
     \begin{tabular}{ccc}
         \includegraphics[keepaspectratio,scale=0.3]{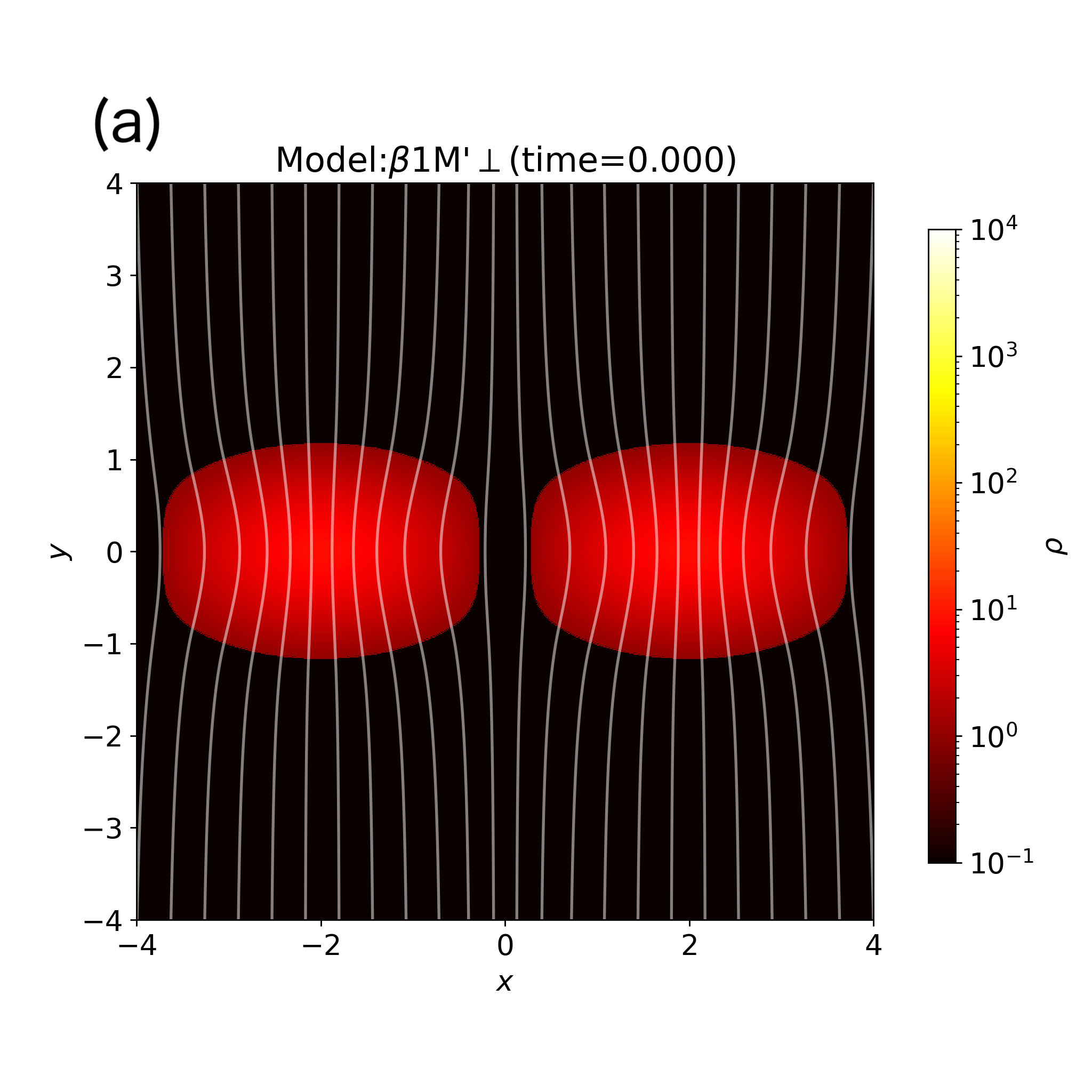}& \includegraphics[keepaspectratio,scale=0.3]{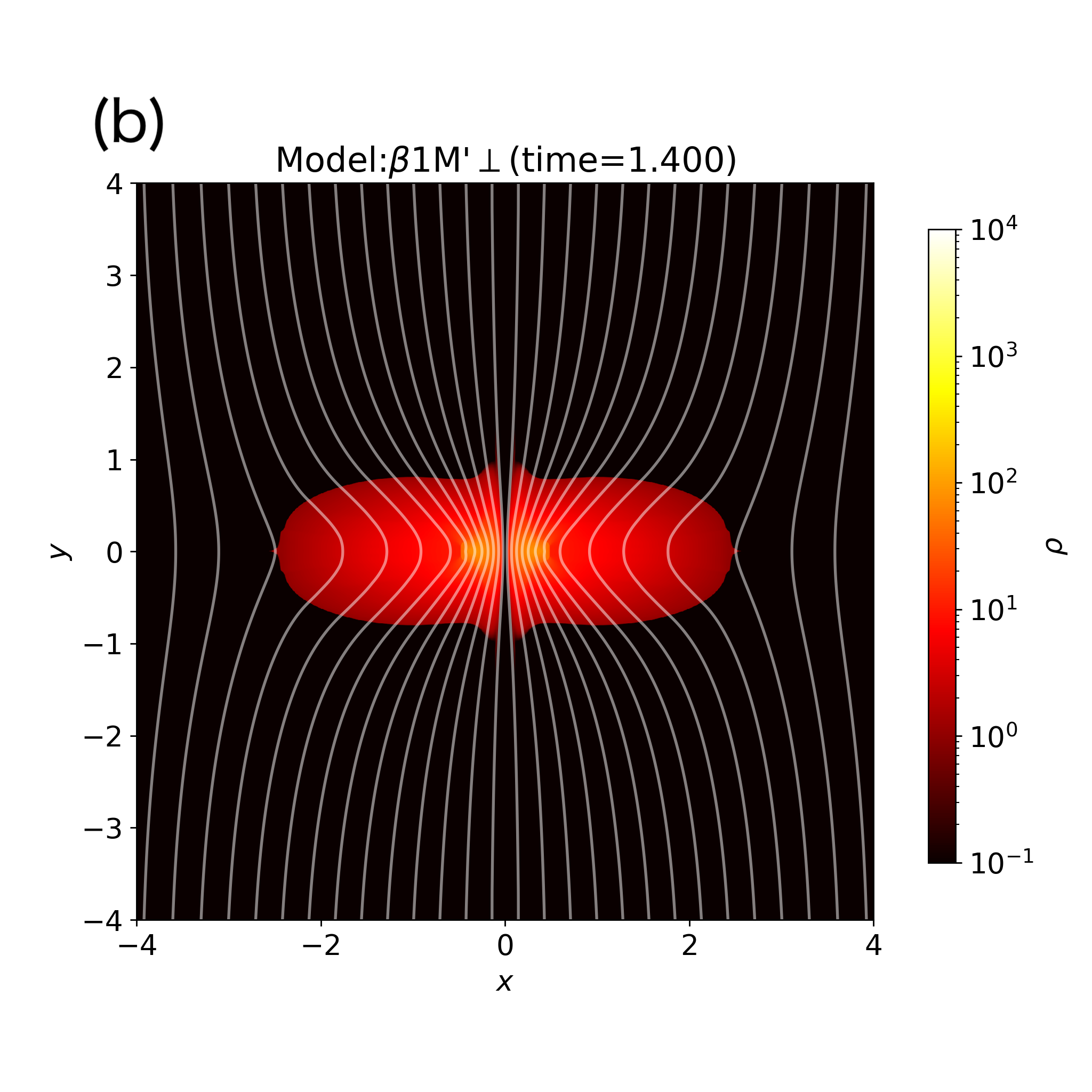} & \includegraphics[keepaspectratio,scale=0.3]{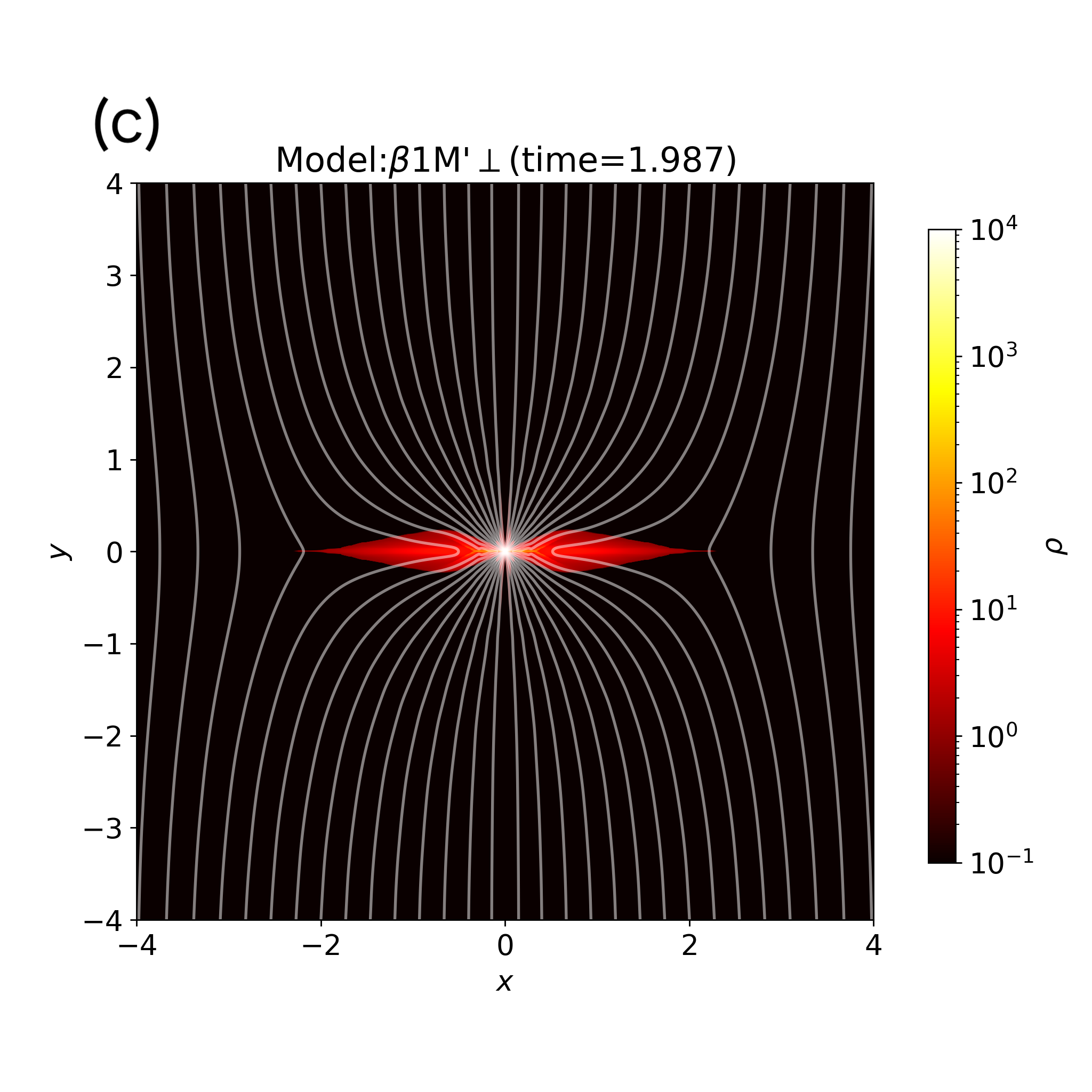}\\ 
         \includegraphics[keepaspectratio,scale=0.3]{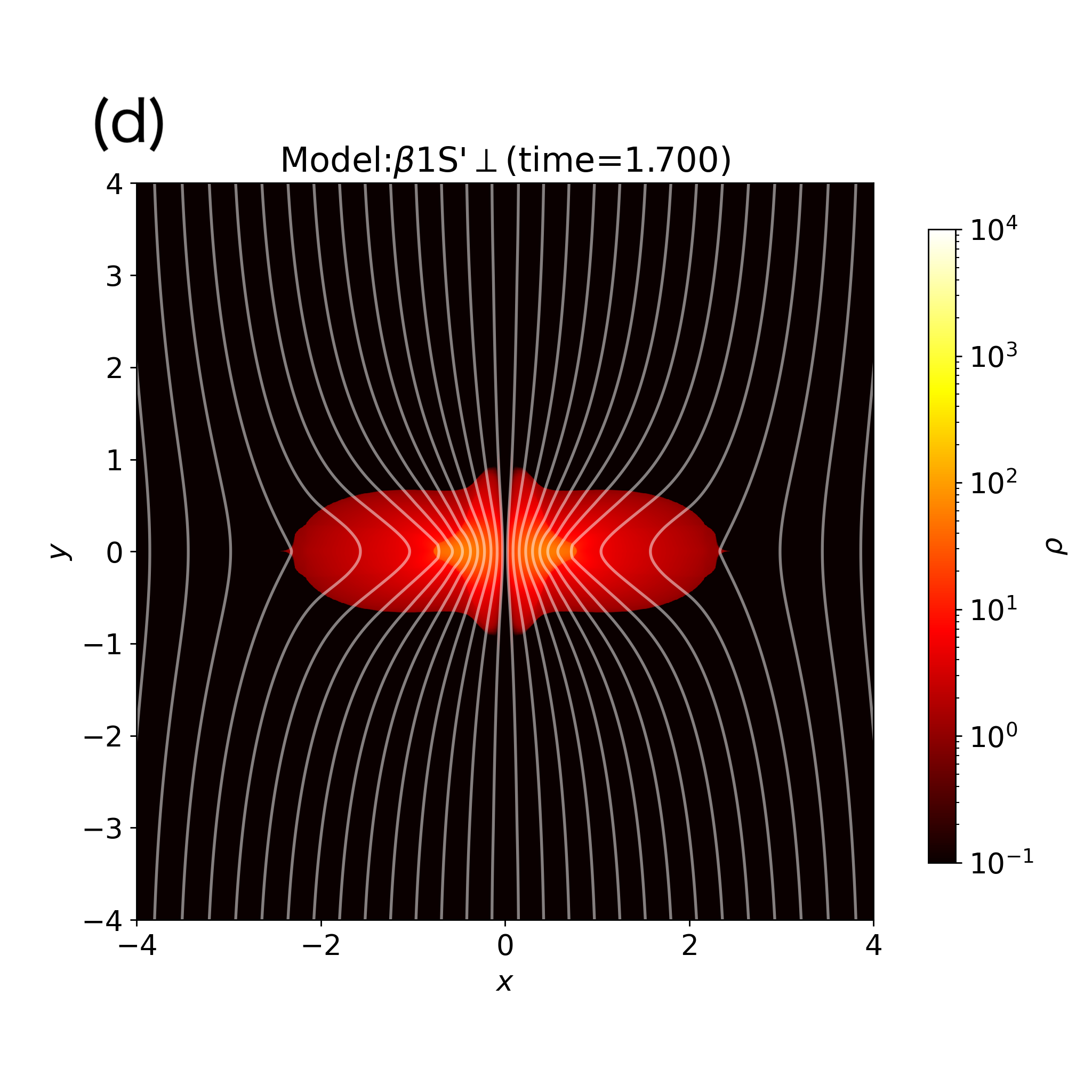}& \includegraphics[keepaspectratio,scale=0.3]{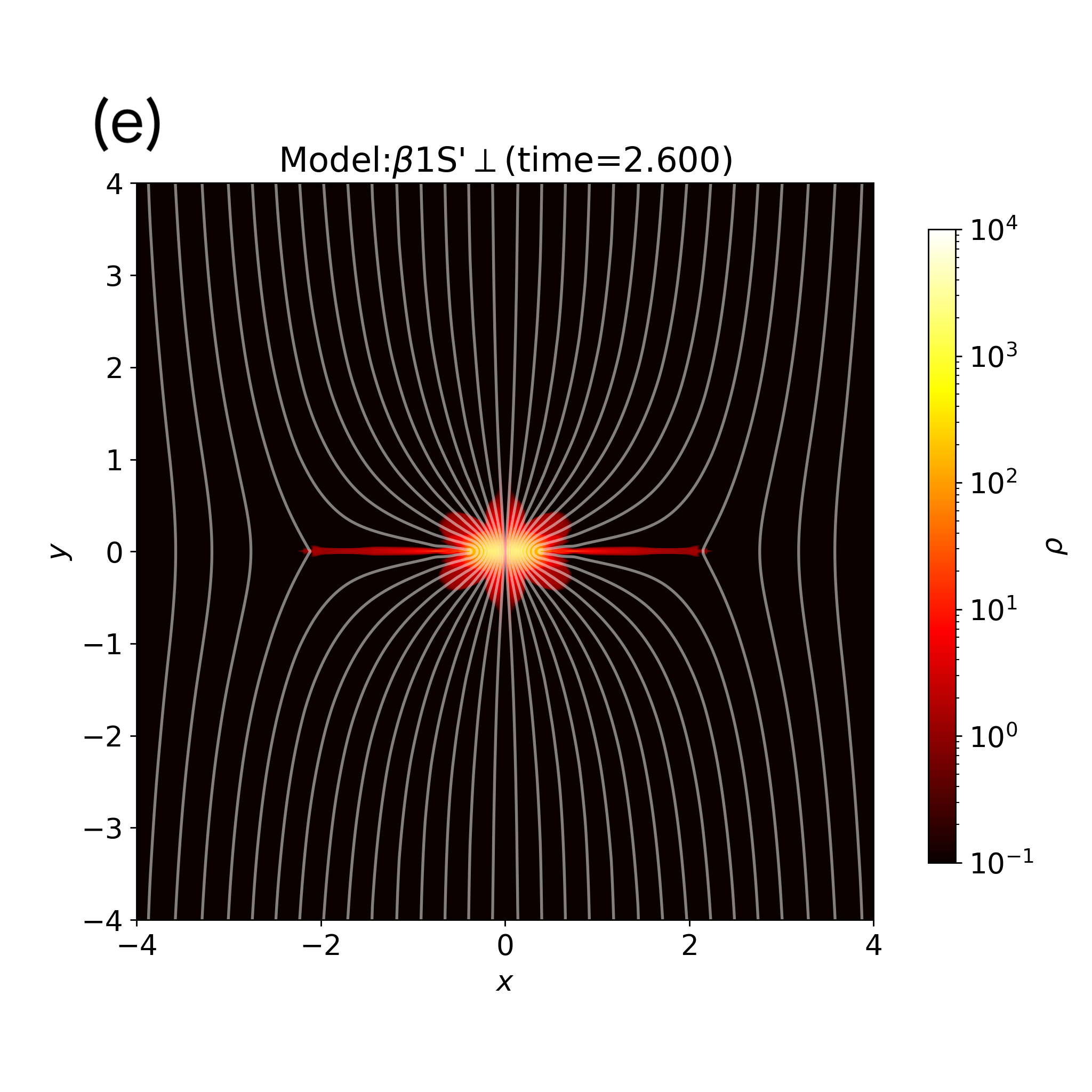} & \includegraphics[keepaspectratio,scale=0.3]{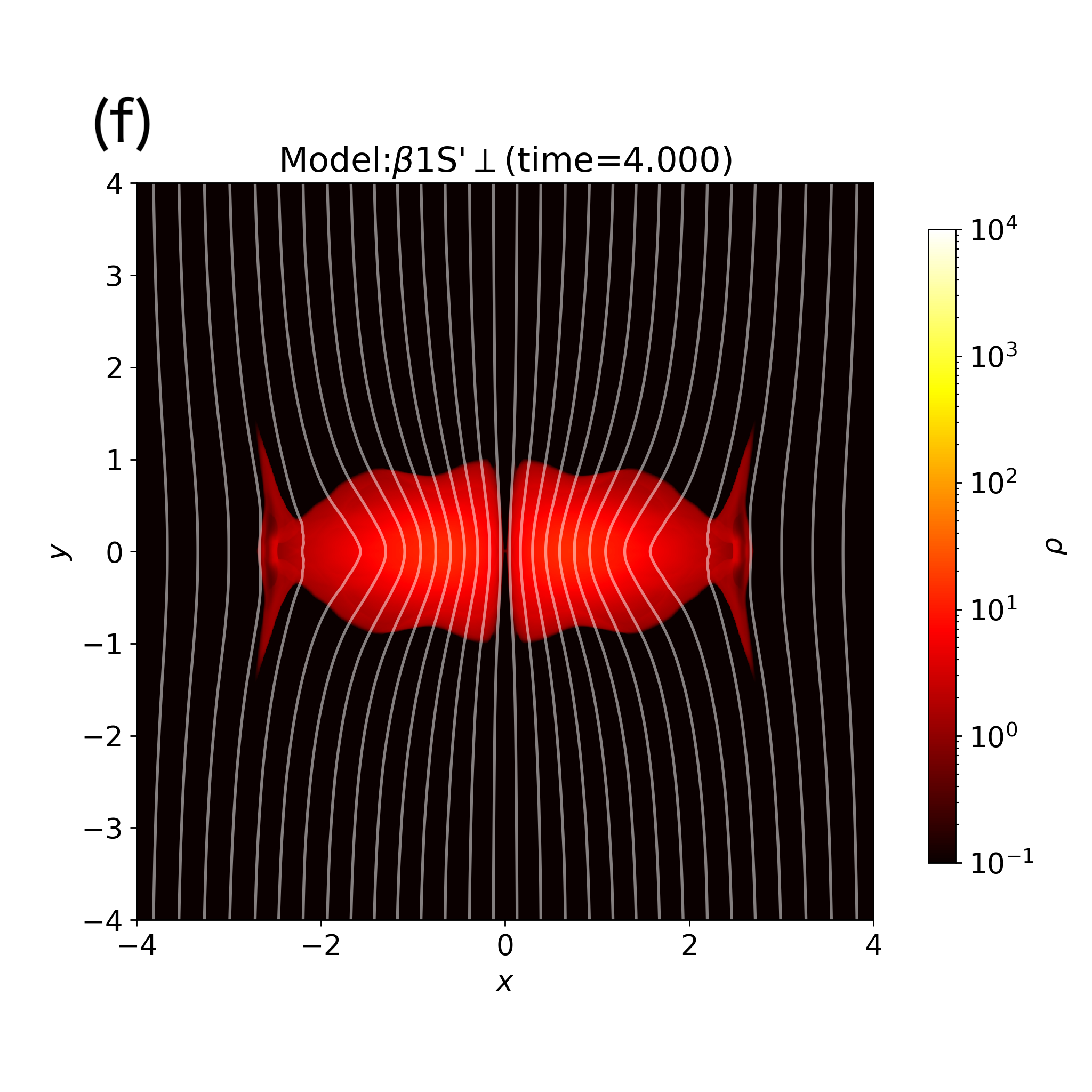}\\  
     \end{tabular}
      \caption{Same as Figure \ref{fig:2d_L075}, but representing the models in which the collision direction ($x$-direction) is perpendicular to the global magnetic field ($y$-direction).
      This figure shows the cross-section of the colliding filaments on the $x-y$ plane, and the plot area is restricted to $\pm4.0$ from the box size ($\pm5$).
      The upper (\textit{a})-(\textit{c}) and bottom (\textit{d})-(\textit{f}) panels show the models of $\beta$1M'$\perp$ and $\beta$1S'$\perp$, respectively. 
      }
     \label{fig:perp_2d}
\end{figure*}

In this section, we investigate the case where the collision direction of the filamentary molecular clouds is perpendicular to the direction of magnetic field lines.
We studied the models in which the magnetic field lines ran parallel to the colliding direction. 
In reality, this happens in a case where the magnetic field lines run perpendicular to the colliding direction.

Figure \ref{fig:perp_2d} shows the time evolution of models $\beta1$M'$\perp$ (upper) and $\beta1$S'$\perp$ (lower), in which the total line mass is considered as $\lambda_\mathrm{tot}=1.64\lambda_\mathrm{crit}$ and $1.40\lambda_\mathrm{crit}$.
The other parameters are set as $\beta_0=1$ and $V_{\rm int}=1$.

In the model of $\beta$1M'$\perp$, the shocked region begins to collapse, whereas, in the model of $\beta$1S'$\perp$, the shocked region does not collapse and is stable while oscillating.
As the filament with $\beta_0=1$ has a magnetically critical line mass of $\lambda_\mathrm{crit,B}=1.17\lambda_\mathrm{crit}$ (Equation (\ref{eq:dimensionless_magnetized_critical_line_mass})), the total line mass of the two models exceeds the critical line mass as $1.4$ ($\beta$1M'$\perp$) and $1.2$ ($\beta$1S'$\perp$) times $\lambda_\mathrm{crit,B}$.
Thus, if a collision occurs in the parallel direction to the magnetic field, both models would undergo the radial collapse (sec. \ref{subsection:criteria}) as $\lambda_\mathrm{tot}>\lambda_\mathrm{crit,B}$.
However, model $\beta1$S'$\perp$ results in a stable oscillation (Figure \ref{fig:perp_2d} lower panels).

This difference is attributed to the collision direction.
In this case, the magnetic flux of the merged filament become twice as large as that in the previous models ($\beta1$S and $\beta1$M), in which the colliding direction is parallel to the magnetic field lines.
Thus, the expression of the threshold line mass should be modified by the magnetic field’s geometry.
When we consider that the magnetic field lines run perpendicular to the collision direction, the magnetically critical line mass (Equation (\ref{eq:magnetized_critical_line_mass})) of the merged filament is changed as
\begin{equation}\label{eq:magnetized_critical_line_mass_perp}
    \lambda_{\rm crit,B}\simeq 0.24 \frac{2\Phi}{G^{1/2}}+1.66\frac{c_s^2}{G}.
\end{equation}
We rewrite the above equation in the normalized form as, 
\begin{equation}\label{eq:dimensionless_magnetized_critical_line_mass_perp}
    \lambda'_{\rm crit,B}\simeq 6.08 \Phi'+20.8.
\end{equation} 
By adopting this equation, the critical line mass of a merged filament is estimated as,  $\lambda_\mathrm{crit,B}=1.51\lambda_\mathrm{crit}$ instead of $1.17\lambda_\mathrm{crit}$.
Therefore, the total line mass of $\beta$1M'$\perp$ model $\lambda_\mathrm{tot}=1.64\lambda_\mathrm{crit}$ exceeds the critical value $\lambda_\mathrm{crit,B}=1.51\lambda_\mathrm{crit}$ and the collapse of the shocked region, even in the perpendicular collision.
On the contrary, for the model $\beta$1S'$\perp$, the total line mass $\lambda_\mathrm{tot}=1.17$ is significantly smaller than the critical line mass $1.51\lambda_\mathrm{crit}$.
Thus, the merged filament of model $\beta$1S'$\perp$ is magnetically subcritical $\lambda_\mathrm{tot} < \lambda_\mathrm{crit,B}$ and exhibits stable oscillations.
Thus, we conclude that the stability of the perpendicular model can be explained by considering whether the merged filament is larger than the magnetically critical line mass if we consider the fact that the magnetic flux increases twice due to the collision.

\section{Summary and Conclusions}\label{sec:summary}
In this study, we performed two-dimensional MHD simulations of head-on collisions between two identical isothermal filaments in magnetohydrostatic equilibrium, in which the magnetic field lines are lateral.
Our findings are summarized as follows:
\begin{enumerate}
\item We studied 14 models in which two filaments shared the same magnetic flux tube, and the collision direction was parallel to the global magnetic field. 
Supersonic collision induced shock waves that swept the merged filament.
A high-density sheet was formed between the shock fronts, which was formed from shock compression and accretion flow through self-gravity. 
The outcome of the shocked high-density sheet was either a radial collapse, in which the central density increased infinitely within a finite time scale, or a stable oscillation with a finite amplitude as a whole.
\item The condition of whether the merged filament undergoes radial collapse or stable oscillation was determined by the initial total line mass of the colliding filaments. When we observed the occurrence of radial collapse after the collision, the initial total line mass ($\lambda_\mathrm{tot}$) exceeded the critical line mass for a magnetically supported filament $\lambda_{\rm crit,B}$, Equation (\ref{eq:magnetized_critical_line_mass}) \citep{2014ApJ...785...24T}.
This condition appeared not to depend on the collision speed.

\item When $\lambda_\mathrm{tot}\leq\lambda_\mathrm{crit,B}$, the merged filament continued to oscillate.
The image of the shocked region resembles the magnetohydrostatic equilibrium state, regardless of the initial relative velocity and the magnetic field strength.

\item In the collapsing model $ \lambda_{\rm tot}>\lambda_{\rm crit,B}$, the density profile of the dense part of the shocked region was affected by the Lorentz force. 
Although the density distribution parallel to the magnetic field remains similar to a sheet-like structure regardless of the magnetic field strength, the density profile became broader in the direction perpendicular to the magnetic field lines as the magnetic field strength increased.
Contrary to this, the density profile of the dense part was not affected by the initial relative velocity, although the distribution changes with the direction of the magnetic field.

\item In the collapsing model $\lambda_\mathrm{ tot}>\lambda_\mathrm{crit,B}$, the line mass of the shocked region $\lambda_{\rho_c/100}$ decreased with time and converged to a certain value, which appears to be independent of the initial line mass.
In the collapsing model, the collapse proceeded in a self-similar manner (Section \ref{sec:self-similar}).

\item 
We examined the timescale of the fragmentation of the merged filament compared with the time scale of the maximum growth.
Our results indicate that stable models may fragment during oscillation maxima or minima, whereas radial collapse models did not provide sufficient time for fragmentation to occur.

\item We studied two models in which the magnetic field lines ran in a direction perpendicular to the collision.
In this case, the radial stability of the shocked region appeared to be given by the same condition, although we must consider the fact that the merged filament had a twice larger magnetic flux than the parallel collisions.

\end{enumerate}

In this study, we tried to make the filament magnetically supercritical as a result of the filament collisions. 
However, we note that there are several mechanisms that can trigger the radial collapse in filament systems.
For example, ambipolar diffusion causes the filaments to collapse radially, even if they are initially subcritical.
Over approximately $10t_\mathrm{ff}$, the ambipolar diffusion leads to increase the mass-to-flux ratio and to decrease in the critical line mass, ultimately resulting in radial collapse, even in stable models. 
Furthermore, gas accretion from the surrounding gas onto filaments works as another mechanism to make filaments supercritical. 
Although there are other possible mechanisms for increasing the filament line mass and initiating star formation other than collisions, the observational fact that active star formation occurs at the hub, the overlapping part of the two filaments, indicates the importance of filament collisions.


\acknowledgements{} 
We would like to express our sincere gratitude to the anonymous referee for their valuable comments and suggestions to improve this paper.
Then, we are grateful to the following scholars for their valuable discussions regarding this research: Kumar, M. S. N., Arzoumanian, D., Hanawa, T., and Takiwaki, T.
This study was supported by JST, the establishment of university fellowships towards the creation of science technology innovation, Grant Number JPMJFS2136, and by JSPS KAKENHI (Grant Numbers: 22J11106(RK), 19K03929, 19H01938 (KI), and 19K03919(KT)).
Numerical computations were performed in Cray XC50 at the Center for Computational Astrophysics, National Astronomical Observatory of Japan.


\bibliography{ref_ffc}{}
\bibliographystyle{aasjournal}   


%



\end{document}